\documentclass[english,authoryear, 12, preprint]{elsarticle}
\usepackage[T1]{fontenc}
\usepackage[latin9]{inputenc}
\usepackage{geometry}
\usepackage{xcolor}
\usepackage{pdfcolmk}
\usepackage{array}
\usepackage{rotfloat}
\usepackage{units}
\usepackage{textcomp}
\usepackage{multirow}
\usepackage{amsmath}
\usepackage{graphicx}
\usepackage{setspace}
\PassOptionsToPackage{normalem}{ulem}
\usepackage{ulem}
\makeatletter

\providecommand{\tabularnewline}{\\}
\providecolor{lyxadded}{rgb}{0,0,1}
\providecolor{lyxdeleted}{rgb}{1,0,0}

\DeclareRobustCommand{\lyxsout}[1]{\ifx\\#1\else\sout{#1}\fi}

\@ifundefined{date}{}{\date{}}
\usepackage{hyperref}
\usepackage{lineno}
\usepackage{graphicx}
\usepackage[english]{babel}

\makeatother

\usepackage{babel}
\begin{document}

\begin{frontmatter}{}

\title{Constraints on effusive cryovolcanic eruptions on Europa using topography
obtained from Galileo images}

\author{Elodie Lesage$^{1}$, Fr\'ed\'eric Schmidt$^{1}$, Fran\c{c}ois
Andrieu$^{1}$, H\'el\`ene Massol$^{1}$}

\address{$^{1}$ Université Paris-Saclay, CNRS, GEOPS, 91405, Orsay, France}
\begin{keyword}
Europa, Icy moon, cryovolcanism, DEM, shape-from-shading, uncertainties
\end{keyword}

\end{frontmatter}{}

\section*{Abstract}

Images of Europa's surface taken by the Galileo Solid State Imager
(SSI) show smooth features measuring a few kilometers, potentially
resulting from eruptions of low-viscosity material such as liquid
cryomagma. We estimated the volume of four of these smooth features
by producing Digital Elevation Models (DEMs) of four Galileo/SSI images.
We used the shape-from-shading technique with special care to estimate
the uncertainties on the produced DEMs and estimated feature volumes
to be between (5.7$\pm$0.9)$\times10^{7}$ m$^{3}$ and (2.7$\pm$0.4)$\times10^{8}$
m$^{3}$. We discussed the implications for putative sub-surface liquid
reservoir dimensions in the case of eruptions induced from freezing
reservoirs. Our previous cryovolcanic eruption model was improved
by considering a cycle of cryomagma freezing and effusion and by estimating
the vaporized cryolava fraction once cryolava spreads onto Europa's
surface. Our results show that the cryomagma reservoirs would have
to be relatively large to generate these smooth features (1 to 100
km$^{3}$ if the flow features result from a single eruption, and
0.4 to 60 km$^{3}$ for the full lifetime of a reservoir generating
cyclic eruptions). The two future missions JUICE (ESA) and Europa
Clipper (NASA) should reach Europa during the late 2020s. They shall
give more information on those putative cryovolcanic regions which
appear as interesting targets that could provide a better understanding
of the material exchanges between the surface, sub-surface and ocean
of Europa.

\section{Introduction}

The Jovian moon Europa is believed to hide a global liquid water ocean
under its ice crust \citep{Khurana_MagneticFieldOcean_1998,Pappalardo_OceanEvidences_1999}.
This ocean is predicted to be in contact with a silicate mantle, which
could allow the chemical exchanges needed to create a rich habitable
environment \citep{Kargel_BrineVolcanism_1991,Kargel_OceanAndCrustComposition_2000}.
The habitability of Europa's ocean depends on chemical conditions
and equilibrium in it \citep{Vance_PhysicalPropertiesEuropa_2018},
but for now, it remains impossible to directly sample. Also, water
coming directly from the ocean seems unlikely to erupt at its surface
because of the very high pressure required for it to ascend through
the whole ice crust \citep{Gaidos_RidgesMeltWater_2000,Manga_PressurizedOceans_2007,Rudolph_FractureIcyShells_2009}.
The strong tides generated by Jupiter in Europa's ice crust could
be at the origin of ice local melting \citep{Tobie_TidalDisspationEuropa_2005,Mitri_TidalPlumes_2008,Vilella_MeltingIcySatellites_2020}.
Such melted reservoirs appear to be good candidates to host life forms
if they were able to remain dormant in the ice between melting episodes
\citep{Gaidos_RidgesMeltWater_2000}. Identifying the geological features
emplaced during eruptions of liquid water could give information on
the location of the terrains that are most likely to show biosignatures,
which can be useful to the two upcoming missions JUICE (ESA) and Europa
Clipper (NASA). In \citet{Lesage_CryomagmaAscent_2020}, we demonstrated
the possibility of erupting liquid water from sub-surface freezing
reservoirs. Here, we propose to test an improved version of our previous
model against Galileo data. First, we select images showing geological
features that could result from the eruption of liquid water/brines
at the surface, i.e. the smooth plains. We then generate the Digital
Elevation Model (DEM) of the chosen features and measure their volumes.
We then use these results to constrain the volume and depth of the
source reservoirs.

The highest resolution images of Europa were acquired during the Galileo
mission with the Solid State Imager (SSI) \citep{Belton_GalileoSSI_1992}.
These images have shown a geologically active surface, characterized
by a wide variety of features \citep{Greeley_GalileoObservations_1998,Greeley_GeologicalMappingEuropa_2000}.
The low crater density found on Europa demonstrates the vigorous resurfacing
processes taking place on this moon, making its surface one of the
youngest in the solar system, with an age under 90 Myrs \citep{Zanhle_CrateringOuterSS_2003}.
Plate tectonics-like processes were thought to frequently recycle
the icy surface \citep{Sullivan_PlateTectonics_1998,Kattenhorn_EvidenceSubduction_2014}
but numerical modeling of the ice shell shows that a global Earth-like
plate tectonics is unlikely on Europa \citep{Howell_PlateTectonicsEuropa_2019}.
Cryovolcanic activity may therefore contribute covering some older
terrains with fresh material. 

A wide range of local-scale geological features is observed at Europa's
surface, such as chaos, lenticulae, domes, pits, and ridges \citep{Greeley_GalileoObservations_1998,Greeley_GeologicalMappingEuropa_2000}.
Several formation mechanisms have been proposed to explain the formation
of these features and invoke, in most cases, a diapiric ascent of
warm ice \citep{Head_RidgesMorphology_1999,Sotin_TidalHeating_2002,Fagents_EuropaCryovolcanism_JGR2003,Schenk_TopographyChaos_2004,Quick_HeatTransfertCryomagma_2016},
or a direct link with the ocean \citep{Greenberg_DiapirChaos_1999,Greenberg_DynamicIcyCrust_2002}.
More recent studies show that the formation of some of these features
could be related to the presence of sub-surface liquid reservoirs,
as for lenticulae \citep{Michaut_DomesAndPitsWater_2014,Manga_Lenticulae_2016},
chaos \citep{Schmidt_ChaosDiapirs_Nature2011} and double ridges \citep{Johnston_CrystallizingWaterBodies_2014}.
Numerical models showed the possibility of generating locally melted
zones within Europa's ice crust due to the combination of convection
and tidal heating \citep{Sotin_TidalHeating_2002,Mitri_TidalPlumes_2008,Han_ConvectionAndTidalDissipation_2010,Vilella_MeltingIcySatellites_2020},
which could explain the formation of sub-surface molten reservoirs.
\citet{Fagents_EuropaCryovolcanism_JGR2003}, \citet{Quick_CryovolcanicDomes_2016},
and \citet{Nunez_DatabaseDomesDEM_2019} suggest that a subset of
domal lenticulae have been put in place by the eruption of cryomagma
reservoirs. Thank to DEMs and numerical modeling, \citet{Quick_CryovolcanicDomes_2016}
demonstrated that domes could by put in place the effusion of viscous
cryomagma. Nevertheless, thinner and smoother features observed on
Galileo images have received less attention.

In their exhaustive classification of Europa's geological features,
\citet{Greeley_GeologicalMappingEuropa_2000} introduced the so-called
``smooth plains'' units, which are defined as smooth surfaces, with
no visible texture, that embay or overprint preexisting terrains.
\citet{Greeley_GeologicalMappingEuropa_2000} proposed two models
of formation for smooth plains, which are (1) the cryovolcanic emplacement
of low-viscosity material (such as liquid water-based mixture) and
(2) the melting of the surface due to a local heat source. Nevertheless,
as we will show it in this study, the smooth plains morphology is not
consistent with surface melting only. Moreover, because of the shape
of the smooth plains, which are very thin and topographically constrained,
these features have been widely interpreted as the result of liquid
flows on the surface \citep{Pappalardo_OceanEvidences_1999,Fagents_EuropaCryovolcanism_JGR2003,Miyamoto_FlowsPatterns_2005}.
\citet{Miyamoto_FlowsPatterns_2005} modeled liquid flow under Europa's
surface conditions and have shown that the effusion of a low viscosity
material such as water or brines may create flow-like features before
freezing that are consistent with the morphologies of some smooth
plains. In this study, we focus on small-scale smooth plains a few
kilometers wide that we call ``smooth features'' hereafter.

In \citet{Lesage_CryomagmaAscent_2020}, we modeled the eruption of
liquid cryolava from a freezing sub-surface reservoir. In the case
of a single eruption, we showed that the erupted volume of cryolava
mainly depends on the reservoir volume and depth. Assuming that smooth
features may be formed by the effusion of liquid cryolava to the surface,
measuring the volume of cryolava flows will provide constraints on
the volume and depth of the cryomagma reservoirs using the framework
and results of \citet{Lesage_CryomagmaAscent_2020}. To estimate the
volume of putative flow features, we generate DEMs of Europa's surface
using the AMES Stereo Pipeline (ASP) \citep{Beyer_ASP_2018}. This
open-source tool has been used previously by several authors to generate
DEMs of Europa's surface: for instance, \citet{Schenk_TopographyChaos_2004}
and \citet{Schmidt_ChaosDiapirs_Nature2011} have generated DEMs of
chaotic terrains; \citet{Dameron_DEMRiges_2015} proposed a statistical
study of double ridge morphology; a large database that includes DEMs
of putative cryolava domes is currently being built by \citet{Nunez_DatabaseDomesDEM_2019}.
Here we develop our own cryovolcanic features database based on two
morphologic criteria (see sec. \ref{subsec:criteria}): (i) a thin
feature confined by the surrounding ridges and (ii) a smooth appearance
with no blocks showing a ridged texture. We selected four images with
smooth features that are compatible with liquid flows, and we calculate
the volume of these features. We also calculate the uncertainty on
the DEMs based on a sensitivity study (see supplementary materials
for details).

\section{Methods}

\subsection{Selection of cryovolcanic features : criteria \label{subsec:criteria}}

\begin{figure}
\centerline{\includegraphics[height=1\textheight]{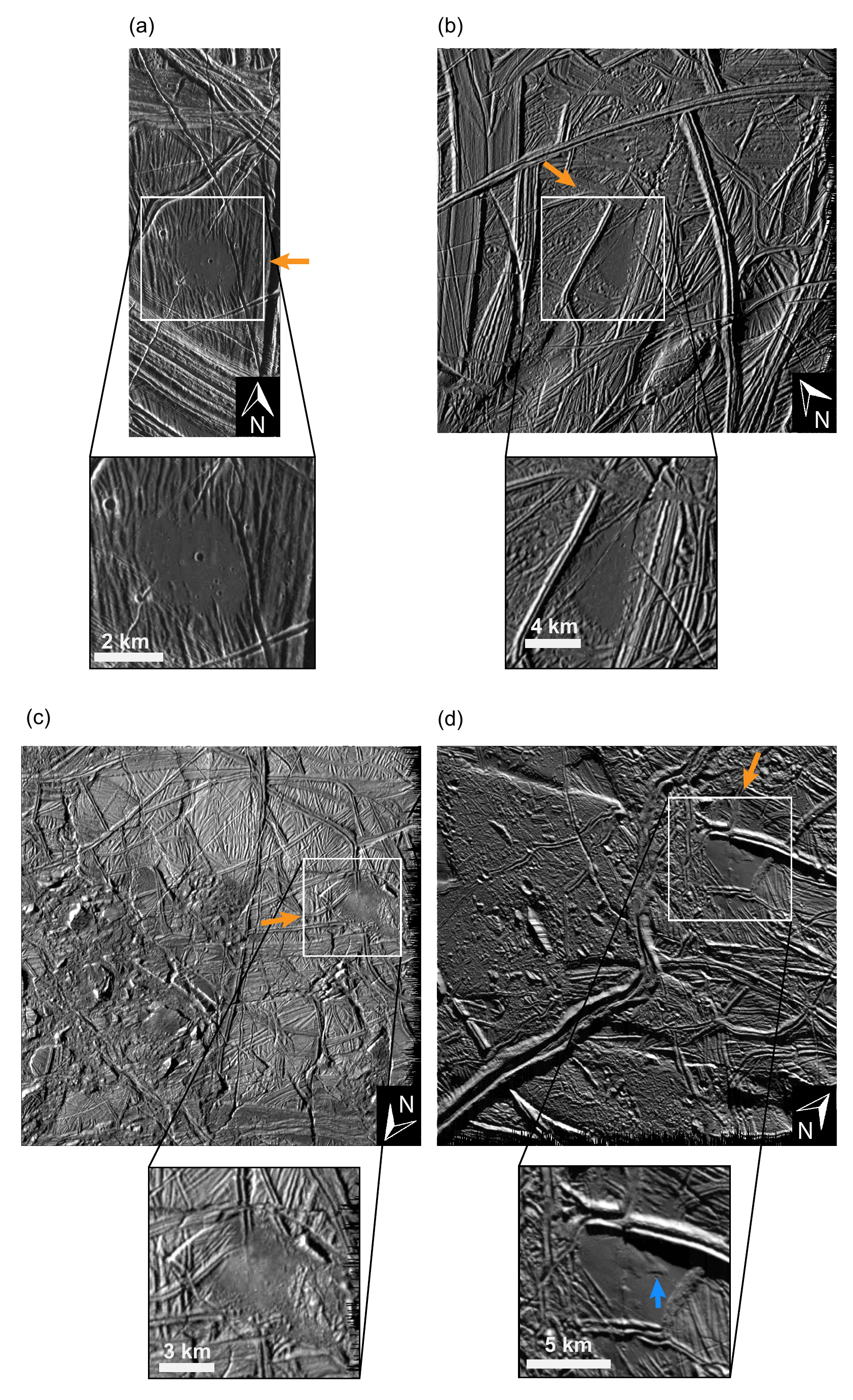}}\caption{Images selected to generate DEMs. (a) Image 5452r (-1°, 340°), resolution:
27 m/px; (b) Image 0713r (-79°, 124°), resolution: 57 m/px; (c) Image
0739r (-81°, 132°), resolution: 57 m/px; (d) Image 9352r (-28°, 218°),
resolution: 60 m/px. All these images are from the Galileo SSI data.
Scales are indicative as these images are not projected. Orange arrows
show the sunlight direction. The blue arrow points out a lobate feature
located on top of the smooth plain of image 9352r. \label{fig:mosaic_images}}
\end{figure}

To obtain the volume of cryovolcanic eruptions putative products,
we generate DEMs of features that might have possibly been put in
place during the effusion of liquid water at the surface. These features,
called hereafter ``smooth features'', are a few kilometers wide
and have common characteristics with smooth plains previously defined
by \citet{Greeley_GeologicalMappingEuropa_2000}. To select the most
relevant features in the Galileo dataset and avoid confusion with
other features, such as chaos, we choose two major criteria to define
the putative cryovolcanic features:
\begin{enumerate}
\item \textbf{Thin features occupying topographic lows. }Because we are
interested in features that may have been emplaced during the effusion
of liquid onto the surface, and as modeled by \citet{Miyamoto_FlowsPatterns_2005}
(``A-type'' features in their study), the resulting features are
expected to be thin, with no particular domal shape. These features
occupy the topographic lows, and are constrained by pre-existing relief.
On Europa, these features are typically delimited by ridges.
\item \textbf{A smooth appearance with no visible blocks of older surface.
}Some features at Europa's surface present a smooth matrix, but also
contain blocks presenting a texture similar to the pre-existing, ridged
plains. These features, named chaos, are identified as the result
of local melting and disruption of the surface, followed by its freezing
\citep{Greenberg_DynamicIcyCrust_2002,Figueredo_MuriasChaos_2002,Schmidt_ChaosDiapirs_Nature2011}.
Here we study smooth features assumed to have been emplaced by effusion
of liquid cryolava onto the surface, which is a completely different
process. Nevertheless, we do not rule out the possibility of forming
new blocks of ice, either during the flowing of water at the surface
because of the very low temperature at Europa's surface, or due to
the transport of small blocks of ice due to stopping as fluid ascends
toward the surface. Hence, we do not exclude features containing relatively
small blocks (covering a surface fraction less than $\sim10\%$ of
the feature), as long as they do not have a ridged surface.
\end{enumerate}
We looked at all SSI images with a resolution higher than $\sim$100
m/px and found 4 regions fulfilling these criteria (see Fig. \ref{fig:mosaic_images}).
A fifth interesting smooth plain was identified (image 8613r) but
unfortunately, we did not manage to produce its DEM because of a large
projected shadow on the image.

\subsection{DEM generation \label{subsec:DEM-generation}}

\begin{figure}
\centerline{\includegraphics[width=1\textwidth]{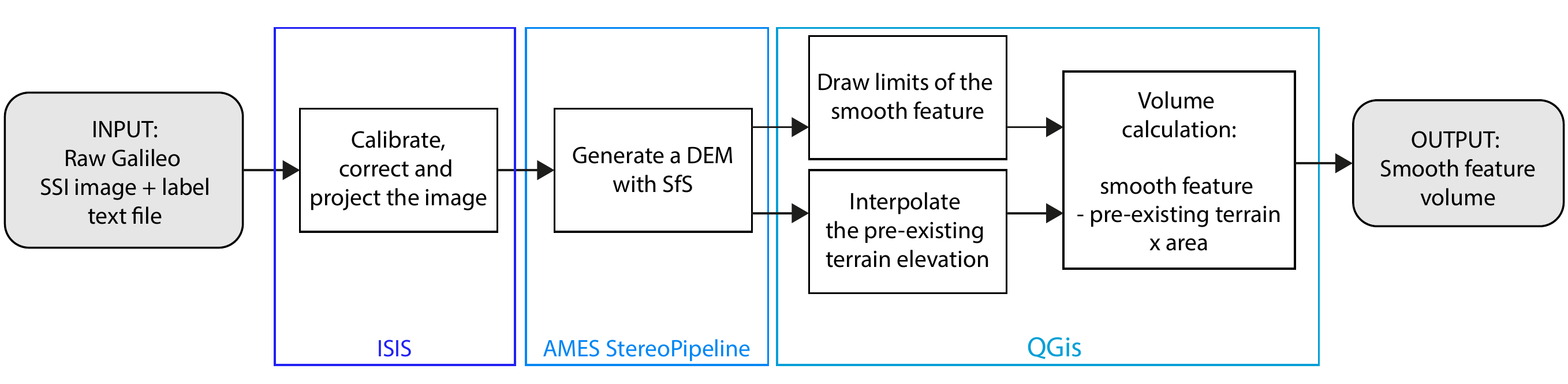}}\caption{Flowchart of the methods used in this study to generate and interpret
DEMs. See supplementary materials for details on the pre-processing
steps using ISIS. DEM generation using the AMES StereoPipeline is
described in section \ref{subsec:DEM-generation}. Post-processing
with QGis and volume calculation are detailed in section \ref{subsec:Gis}.
\label{fig:flowchart_method} }
\end{figure}

A few steps are required to obtain a ready-to-use DEMs . After calibration,
noise-filtering, and map-projection of the images using ISIS 3 (https://isis.astrogeology.usgs.gov,
see details in supplementary materials, section 1 and in the ``ISIS''
part of flowchart in Fig. \ref{fig:flowchart_method}), we use the
NASA AMES StereoPipeline to generate the DEMs (\citealp{Beyer_ASP_2018},
``AMES StereoPipeline'' part of the flowchart in Fig. \ref{fig:flowchart_method}).
These DEMs are then used to calculate the smooth feature volumes (``QGis''
part of the flowchart in Fig. \ref{fig:flowchart_method}). 

Two main tools can be used with ASP in order to produce DEMs: the
Stereo tool, based on the stereoscopic analysis, and the Shape from
Shading (SfS) tool, based on the photoclinometric principle. The Stereo
tool produces robust DEMs because this method is based on the correspondences
between pixels of two (or more) images \citep{Beyer_ASP_2018}. Stereoscopy
requires at least two images of the same terrain, each taken from
a different point of view (at least a few degrees of difference) and
illuminated from a similar direction. These two criteria are very
limiting for the use of the Galileo SSI data: due to the limited number
of high-resolution images, image pairs that satisfy these two criteria
are extremely rare. For this reason, we cannot use the Stereo tool
to produce DEMs of the smooth plains, and we focus on the SfS. The
ASP Stereo tool is expected to give more robust results, but SfS produces
3 to 5 times better resolved DEMs \citep{Nimmo_StereoSfSCompare_2008},
which allows a more precise study of small scale elevation variations.
Also, even though SfS is only able to give relative heights and cannot
be used to infer an absolute elevation, this is not limiting in our
case as we only want to know the height of cryovolcanic features relative
to the surrounding terrain. 

To produce the DEMs presented here, we use the SfS tool based on the
photoclinometric principle \citep{Alexandrov_SfsASP_2018}. SfS uses
light intensity of each pixel of an image to infer the surface topography.
In fact, the mean brightness of each surface facet (i.e. the footprint
of each pixel on the real surface) is a function of the angle between
the sunlight incidence direction and the direction normal to the facet.
Based on a known solar elevation and azimuth, SfS computes the slope
of each image pixel from its brightness. Then, it integrates pixel
slopes to give the terrain shape. Numerically, this process is done
by minimizing the following cost function \citep{Alexandrov_SfsASP_2018}:
\begin{equation}
\int\int\left[I\left(h\left(x,y\right)\right)-T.A.R\left(h\left(x,y\right)\right)\right]^{2}+\mu^{2}\parallel\nabla^{2}h\left(x,y\right)\parallel^{2}+\lambda^{2}\left[h\left(x,y\right)-h_{0}\left(x,y\right)\right]^{2}dx\:dy\label{eq:cost_func}
\end{equation}

where $h(x,y)$ is the function describing the terrain topography,
$I\left(h(x,y)\right)$ is the image viewed by the camera reconstructed
using the terrain $h(x,y)$, $T$ is the image exposure, $A$ is the
terrain albedo (considered to be constant on the whole image) and
$R\left(h\left(x,y\right)\right)$ is the reflectance. $h_{0}\left(x,y\right)$
is an ``initial guess'' terrain (also called ``apriori'' in the
SfS tool) from which the minimization algorithm starts to calculate
the real shape of the terrain. This initial guess is provided by the
user in the form of a geolocalized DEM. The best initial guess would
be a DEM generated with the stereoscopic technique, which could be
refined by the SfS tool. Here we do not possess such a DEM, so we
provide a flat DEM to SfS to start the minimization. Finally, $\mu$
and $\lambda$ are two positive coefficients chosen by the user and
controlling respectively the smoothing of the DEM and the weight of
the initial guess terrain. 

The first term of the cost function constrains the brightness and
ensures that the simulated light intensity fits with the image recorded
by the camera. This term depends on the reflectance model used, which
can be chosen by the user. The Lambertian, LunarLambert and Hapke
models are available to model icy surfaces. In accordance with a recent
photometric study \citep{Belgacem_RegionalPhotometryEuropa_2019},
we use the Hapke model with the following parameters: $\omega$=0.9,
$b$=0.35, $c$=0.65, $B_{0}$=0.5, $h$=0.6. The influence of the
photometry on the DEMs is showed in supplementary materials (section
2). The relative uncertainty due to photometry has been estimated
at $\pm$10\% of the measured volume (see supplementary materials).

The second term of the cost function controls the smoothness of the
output DEM by minimizing the second-order derivative of the slope
on each point. The smoothness coefficient $\mu$ is chosen by the
user. Theoretically, the higher $\mu$, the smoother the DEM, making
the small-scale details less visible and flattening higher relief
features. Nevertheless, we noticed that very small values of $\mu$
also produce flattened DEMs. To avoid an extremely flattened DEM,
we typically choose 1<$\mu$<10. We tested several values of this
coefficient for each image and concluded that there is no ideal value
of $\mu$ that can be used for all the images. In fact, the smoothing
effect controlled by $\mu$ depends on the terrain roughness and therefore
differs for each studied image \citep{Alexandrov_SfsASP_2018}. For
each image, we need to test several $\mu$ values to keep the most
appropriate one. The specific effects of $\mu$ variation on the produced
DEM are shown in the supplementary materials (section 2). For small
features, we found that the DEM is always consistent with the feature
heights estimated from shadows (see supplementary materials section
2), in a large domain of the smoothness parameter $\mu$. The relative
uncertainty due to the smoothness coefficient has been estimated at
$\pm$5\% of the measured volume.

Finally, the third term of the cost function describes the difference
between the calculated DEM and the initial guess $h_{0}(x,y)$ given
to SfS. Here, we do not have an initial guess of the terrain elevation
and SfS is the only tool used to generate the DEMs. Hence, we use
a flat terrain at 0 elevation as the initial guess, and we set the
parameter $\lambda$=0. By doing so, SfS should not depend on this
flat initialization DEM during the minimization iterations.

Finally, the total relative uncertainty of the volumes measured on
the DEMs is $\pm$15\% (see supplementary materials).

\subsection{Volume estimation \label{subsec:Volumes-estimation}}

In order to estimate the cryolava volume erupted during smooth plain
emplacement, the first step is to calculate the total smooth feature
volume from the DEM. This process is detailed in the next section.
With the example of image 5452r, we show in section \ref{subsec:5452r}
that it is necessary to invoke subsidence to explain the morphology
of the smooth features. To obtain a more realistic cryolava volume,
we also take into account the estimated underlying pre-existing topography
(ridges in our case, coupled with terrain subsidence). The hypothesis
and methods used to estimate the pre-existing topography are detailed
in section \ref{subsec:alpha}.

\subsubsection{GIS processing: simple approach \label{subsec:Gis}}

We first calculate the smooth feature volume using QGis \citep{QGis}
based on a simple approach. The idea is to subtract the pre-existing
terrain beneath the cryolava flow from the feature itself in order
to obtain the volume of cryolava. A few steps are necessary to obtain
this result and are summarized in the ``QGis'' part of the flowchart
given in Fig. \ref{fig:flowchart_method}. First, the smooth feature
needs to be delimited. We draw the shape of the feature and exclude
ridges that might intersect it. Then, the pre-existing terrain beneath
the cryovolcanic feature is inferred from topography surrounding the
feature. In order to estimate it, we choose several reference points
in the valleys around the smooth feature, in which the cryolava has
flowed, and we use the interpolation tool in QGis to create a layer
that approximates the terrain elevation under the feature. Once the
pre-existing terrain elevation is subtracted from the DEM, we sum
the height of all the pixels composing the feature and multiply it
by the area of a pixel, so we obtain the volume of the cryolava flow.
This volume is called ``measured volume'' (noted $V_{measured}$)
in the following sections.

\subsubsection{Subsidence and thermal erosion: example of image 5452r \label{subsec:5452r}}

Image 5452r was taken by the Galileo spacecraft during its fourth
orbit with a resolution of approximately 30 m/px. It shows a very
smooth circular feature. This feature was first presented by \citet{Head_Cryovolcanism_1998}
and described by \citet{Pappalardo_OceanEvidences_1999} as a ``smooth
deposit, probably emplaced as a cryovolcanic eruption of low-viscosity
material, perhaps liquid water''. This feature interpretation is
now accepted in literature because of its morphology \citep{Fagents_EuropaCryovolcanism_JGR2003,Miyamoto_FlowsPatterns_2005}.

Some impact craters are visible on the smooth feature, especially
a large one, almost centered. We assume that all craters have been
formed after the emplacement of the smooth feature and have no relation
with its formation. The first reason is because other craters are
also present around the smooth feature and do not seem to interact
with the older terrain (no melting, no particular ejecta, etc.), and
the second reason is because if the smooth feature has been formed
by melting after the impact, the impact crater itself would not be
visible.

\begin{figure}
\centerline{\includegraphics[width=1\textwidth]{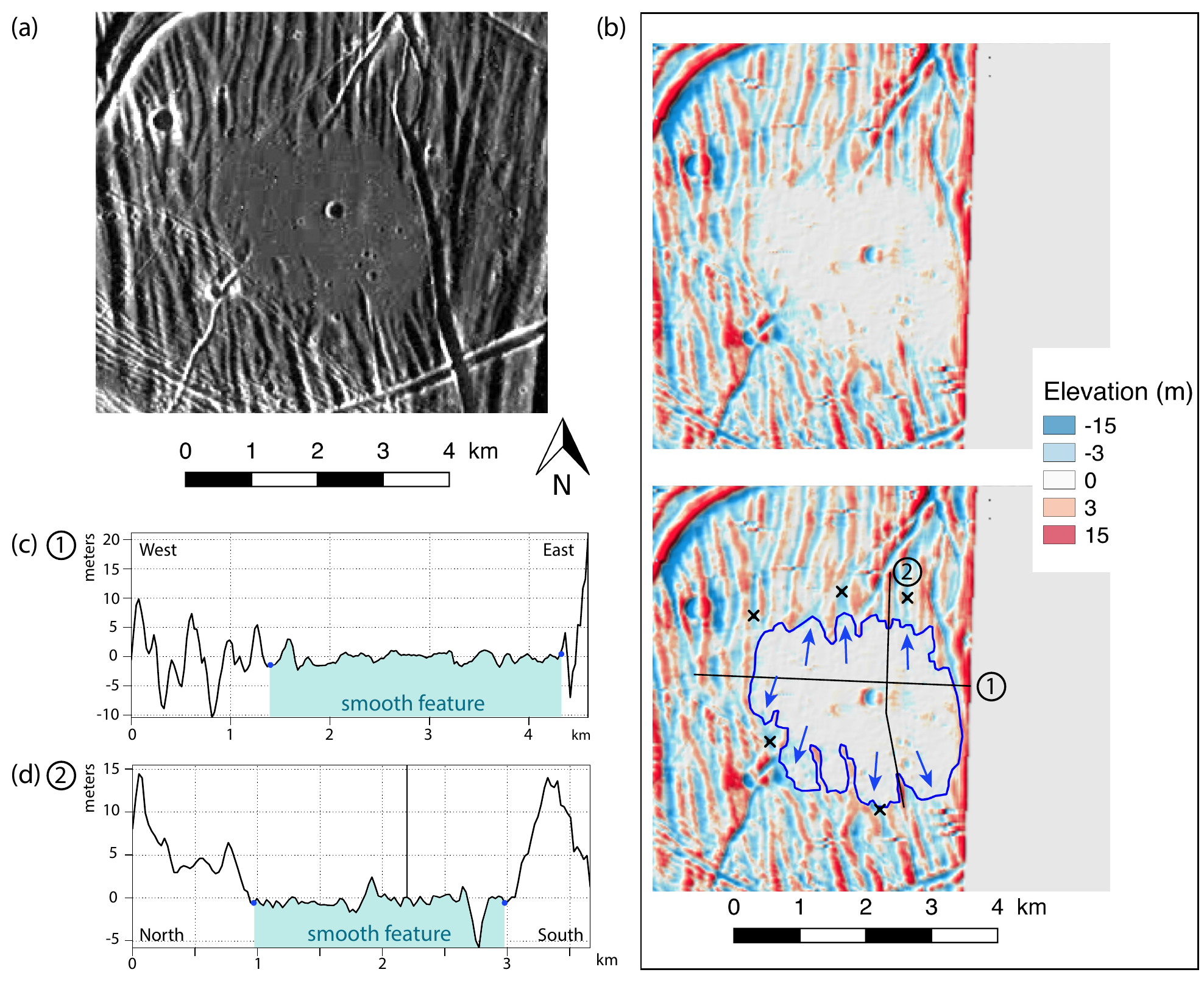}}\caption{(a) Image 5452r from Galileo SSI. Sun light comes from the East (right
on the image).(b) The DEM of image 5452r produced with SfS. The blue
line indicates the limit of the flow-like feature. Blue arrows show
the lobate zones interpreted as filling the valleys between the ridges.
The two profiles numbered (c) 1 and (d) 2 are respectively perpendicular
and parallel to the surrounding ridges. Schematic representations
of feature sections along these two profiles are shown in Fig. \ref{fig:topo_coulee}.
Black crosses indicate the chosen points used to calculate the reference
level of the pre-existing terrain beneath the smooth feature (see
Fig. \ref{fig:topo_coulee}). \label{fig:5452r} }
\end{figure}

\begin{figure}
\centerline{\includegraphics[height=0.75\textheight]{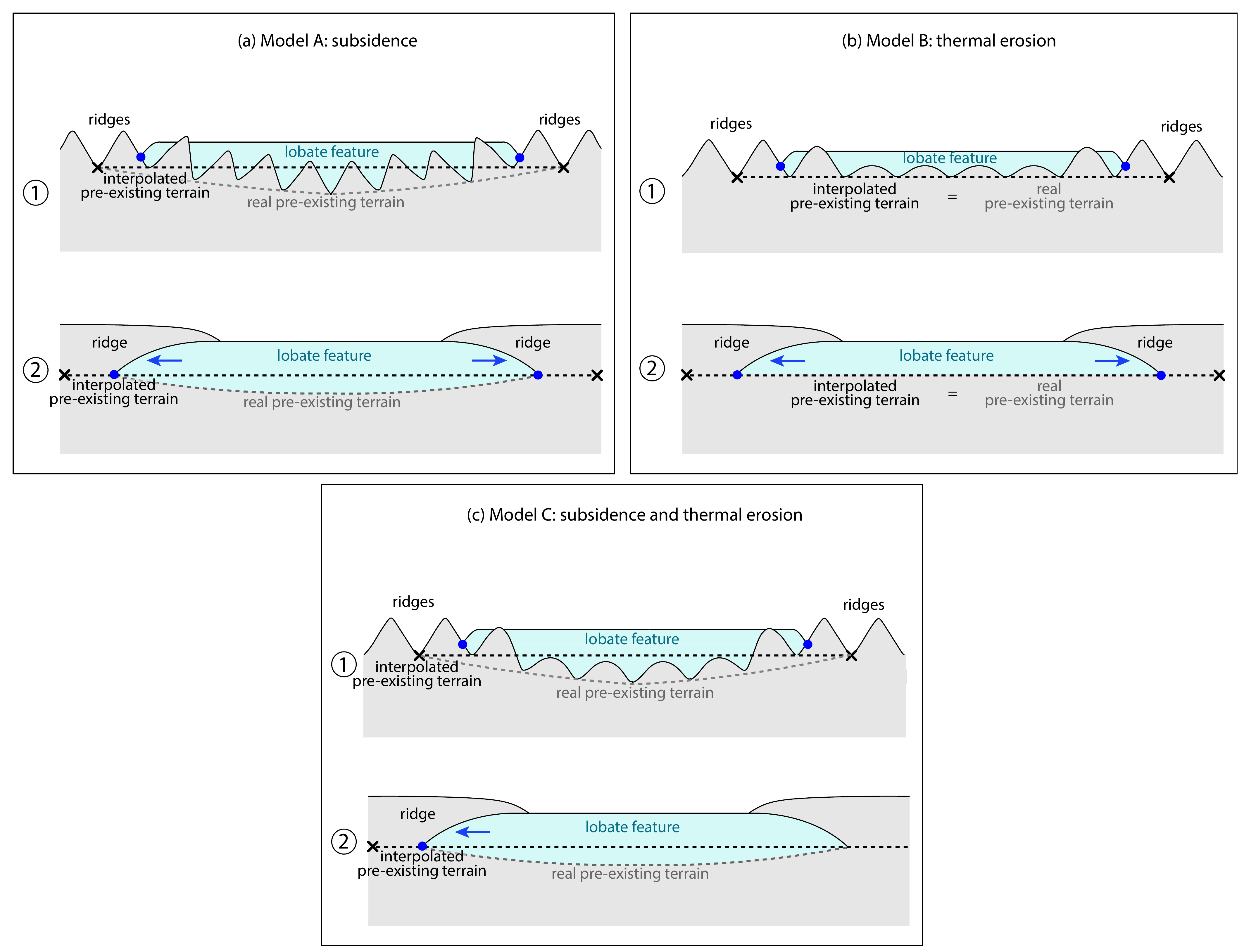}}\caption{Schematic views of the topographic profiles 1 and 2 respectively perpendicular
and parallel to the surrounding ridges (see Fig. \ref{fig:5452r}).
The gray zone represents the pre-existing ridged terrain whereas the
light blue zone stands for the smooth feature. The dark blue points
delimit the edge of the smooth feature corresponding to the dark blue
line in Fig. \ref{fig:5452r}. The black crosses show the bottom of
the nearest ridges which are used as reference elevations to interpolate
the pre-existing terrain mean elevation. This interpolated pre-existing
terrain elevation, represented by the black dotted line, is used to
calculate the feature volume. The actual pre-existing terrain of the
feature may be different from the interpolated one and is plotted
with a dotted gray line. Three mechanisms are sketched: (a) surface
subsidence under the feature (model A), (b) melting or thermal erosion
of the pre-existing terrain (model B) and (c) a mix of these two processes
(model C). A difference between the interpolated pre-existing terrain
and the real one exists in the case of Model A and C. \label{fig:topo_coulee} }
\end{figure}

Image 5452r and its DEM generated with SfS are shown respectively
in Fig. \ref{fig:5452r}a and \ref{fig:5452r}b. We determine the
smooth feature edges (in blue on Fig. \ref{fig:5452r}b) using the
DEM. In fact, in the particular case of image 5452r, the sunlight
comes from the east side of the image, which is the direction perpendicular
to the surrounding ridges and thus makes the east-west oriented lobes
hardly visible on the raw image. Nevertheless, the edges of the feature
are well visible on the DEM, more specifically in the valleys between
the ridges. We also choose five points located in valleys adjacent
to the feature (see the black crosses on Fig. \ref{fig:5452r}b).
These points are used to estimate the pre-existing terrain elevation
beneath the feature (see dashed lines in Fig. \ref{fig:topo_coulee}).
By interpolation, we obtain a surface that is subtracted from the
smooth feature. By doing this, we can calculate the feature volume
itself. DEMs of the three other images used in this study are given
in supplementary materials, section 3.

We propose some geomorphological interpretations of this smooth feature.
First of all, the smooth feature, mostly in white on the DEM, has
an elevation of around 0 m, which is higher than the surrounding valley
bottoms, that have a negative elevation. At the locations indicated
by the blue arrows in Fig. \ref{fig:5452r}, one can see the feature
edges, filling the bottom of the valleys. This is in agreement with
the hypothesis that the smooth feature is possibly made of low-viscosity
material added on the preexisting terrain, flowing between the ridges. 

Moreover, on the north, south, and west of the smooth feature, we
can see that some ridges are not covered by the smooth material, which
indicates that the putative flow had a relatively low viscosity \citep{Miyamoto_FlowsPatterns_2005}.

Finally, at the central region of the smooth feature, ridges are not
visible. This was unexpected because a flow that covers a pre-existing
terrain should theoretically have an elevation higher than the pre-existing
terrain. But here, the surrounding ridges have an elevation higher
than the smooth feature. Two scenarii can be put forward to explain
this effect and are illustrated in Fig. \ref{fig:topo_coulee} (see
the two topographic profiles numbered 1 and 2 in Fig. \ref{fig:5452r}b,
\ref{fig:5452r}c and \ref{fig:5452r}d). The emplaced material could
generate a local subsidence of the ice crust (model A) or it could
melt/erode the surface as it flows (model B). Both cases, or a mix
of them (model C), might explain the lower topography of the pre-existing
terrain especially near the center.

Model A detailed in Fig. \ref{fig:topo_coulee}a could result from
the local subsidence of the ice crust due to the presence of a liquid
reservoir at depth centered on the feature. The required condition
to create a few kilometers of large depletion is a thin elastic layer
of less than a few hundred meters beneath the reservoir \citep{Manga_Lenticulae_2016}.
With this formation model, the real pre-existing terrain beneath the
smooth feature has an elevation lower than the pre-existing terrain
estimated with the DEM, which is the mean level of the surrounding
valleys. This has to be taken into account in the volume measurement
(see section \ref{subsec:alpha}).

In the case of model B, as shown in Fig. \ref{fig:topo_coulee}b,
the ridges are subject to thermal erosion only. This could happen
if a warm liquid flows onto the surface (see thermal erosion experiment
from \citealp{Kerr_ThermalErosionLava_2001}). In this case, the flowing
liquid would be composed of a mixture of cryolava and molten terrain.
After freezing, the molten terrain would return to a frozen state
with a density similar to its original state, so finally, the net
volume change due to thermal erosion would be null. In the case of
image 5452r, the smooth feature is around 0 m elevation, which is
also the mean elevation of the surrounding terrain based on the initial
guess terrain chosen for the DEM generation. This would mean that
in the case of terrain melting and freezing, no material coming from
the interior was added onto the surface and thus only heat transfer
is responsible for the feature emplacement. This is not in agreement
with our volume measurement results (see section \ref{subsec:Measured_and_erupted_volumes}),
so terrain melting only cannot explain the smooth feature morphology,
and we do not consider model B hereafter.

Model C is a combination of thermal erosion and local subsidence,
as illustrated in the sketch from Fig. \ref{fig:topo_coulee}c. As
for model A, the real pre-existing terrain is lower than the estimated
one, implying a difference between the flow volume measured from the
DEM and the real cryolava volume coming from the reservoir. This has
to be taken into account and is detailed hereafter.

\subsubsection{From measured volumes to cryolava volumes \label{subsec:alpha}}

On Europa, most of the surface is ridged (see images in Fig. \ref{fig:mosaic_images}
and \citealp{Greeley_GeologicalMappingEuropa_2000}). Thus, the simple
method proposed in section \ref{subsec:Gis} may overestimate the
volume of actual erupted material as it also takes ridges volume into
account. As we are interested in the actual volume of cryolava erupted
onto the surface, we propose to use a volume factor $\alpha_{V}$
which expresses the actual cryolava volume $V_{cryolava}$ with respect
to the apparent volume $V_{measured}$ measured with the DEMs:
\begin{equation}
V_{cryolava}=\alpha_{V}V_{measured}
\end{equation}

\begin{sidewaysfigure}
\centerline{\includegraphics[width=1\textwidth]{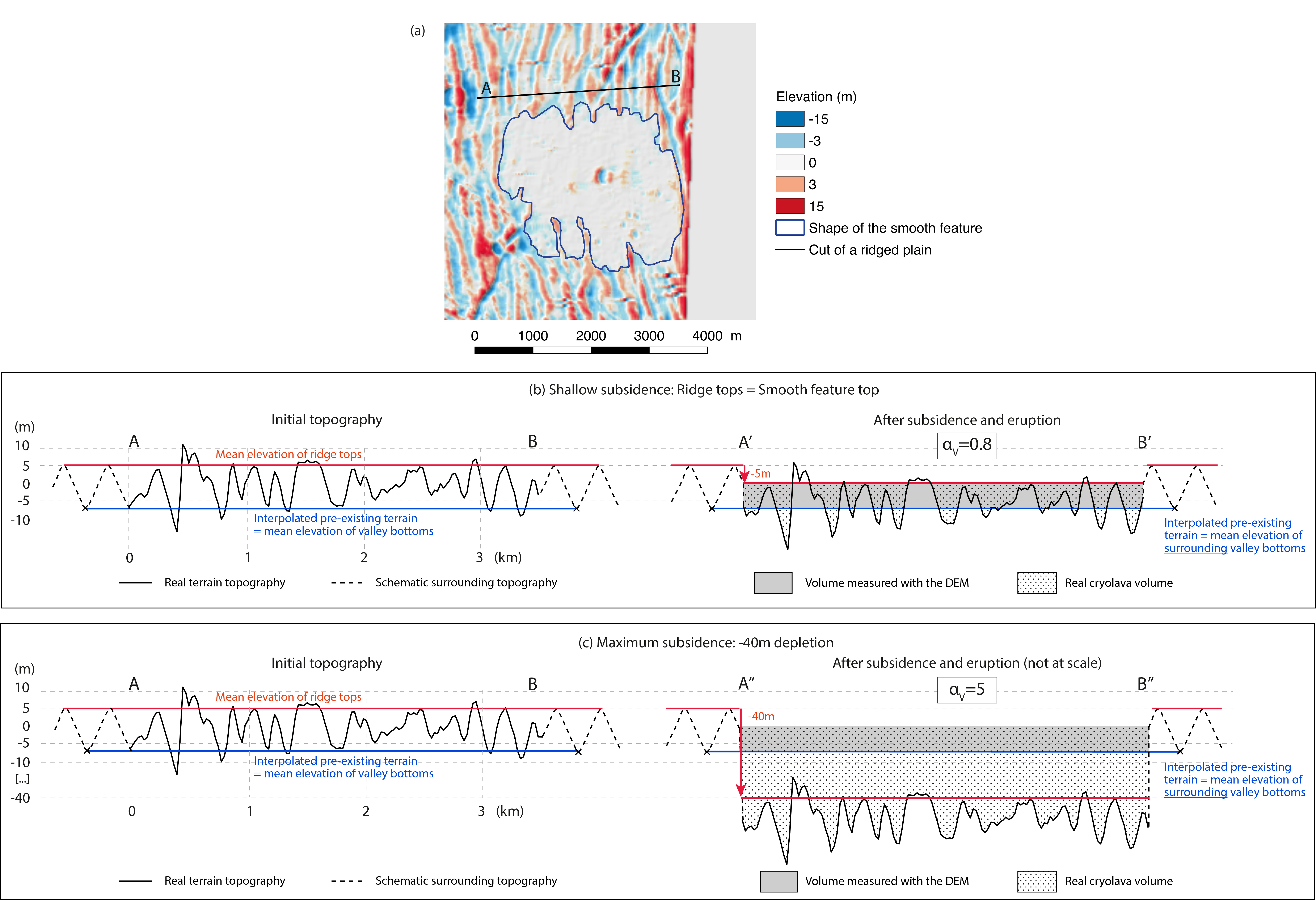}}\caption{Illustration of the calculation of the $\alpha_{V}=\frac{V_{cryolava}}{V_{measured}}$
factor. (a) We make a section in a ridge terrain and we extract a
topographic profile AB from this section. (b and c) We measure the
cross-sectional area of the smooth feature obtained with the simple
approach $A_{measured}$ (in gray), i.e. the difference between the
mean elevation of the smooth feature's top and the elevation of the
interpolated pre-existing terrain multiplied by length AB. We also
measure a more realistic cross-sectional area which should be filled
with cryolava $A_{cryolava}$ (dotted). This cross-sectional area
is defined as the difference between the mean elevation of the smooth
feature's top and the elevation of a more realistic pre-existing terrain,
i.e. a ridged terrain with subsidence of 5 to 40 m, multiplied by
the length AB. We then calculate $\alpha_{V}=A_{cryolava}/A_{measured}$.
We investigate the impact of putative terrain subsidence on $\alpha_{V}$.
We simulate a subsidence of the terrain in 2 cases: (b) a shallow
subsidence, where the ridges tops are showing on the surface (-5 m,
profile A'B'), and (c) a maximum subsidence value after \citet{Manga_Lenticulae_2016}
(-40 m, profile A''B''). \label{fig:topo_ridges}}
\end{sidewaysfigure}

The calculation of $\alpha_{V}$ is illustrated in Fig. \ref{fig:topo_ridges}
and described hereafter. We use a topographic profile AB extracted
from a nearby ridged plain from image 5452r to estimate $\alpha_{V}$
(Fig. \ref{fig:topo_ridges})

In Fig. \ref{fig:topo_ridges} b and c on the left, we show a reference
situation where cryolava flows between the ridges without any subsidence.
We calculate the cross-sectional area of the smooth feature measured
with the simple approach $A_{measured}$, which is the mean height
of the smooth feature top (elevation $\sim$0 m) above the mean elevation
of valley bottoms ($\sim$-7 m) multiplied by the feature's width.
This cross-sectional area is filled in gray in Fig. \ref{fig:topo_ridges}
b and c on the right. 

The situation presented in Fig. \ref{fig:topo_ridges} b and c on
the left might not be realistic because, as discussed above, a cryolava
flow on the surface cannot produce the observed features without local
subsidence (see Fig. \ref{fig:topo_coulee}). Such subsidence may
significantly impact $\alpha_{V}$ factor, as estimated in Fig. \ref{fig:topo_ridges}b
and c on the right. We simulate the subsidence of cross-section AB
to calculate the associated $\alpha_{V}$. We study two extrema to
estimate the range of possible cases: (i) shallow subsidence of only
the necessary height for the ridges to be embayed in the smooth feature
($\sim$5 m, Fig. \ref{fig:topo_ridges}b), (ii) deeper subsidence
of 40 m, which is the maximum subsidence modeled by \citet{Manga_Lenticulae_2016}
for subsurface reservoirs less than 10 km wide (Fig. \ref{fig:topo_ridges}c).
Hence, we calculate the cross-sectional area $A_{cryolava}$, which
is filled with cryolava after the eruption and subduction of the terrain,
represented in dotted in Fig. \ref{fig:topo_ridges} b and c on the
right. We finally calculate $\alpha_{V}=A_{cryolava}/A_{measured}$
using these two areas. This 2D approximation is justified by the homogeneity
of the ridged terrains. Based on cross-sectional areas ratio $A_{cryolava}/A_{measured}$
measured on Fig. \ref{fig:topo_ridges}b and c, we obtain $\alpha_{V}\simeq0.8$
for shallow subsidence and $\alpha_{V}\simeq5$ for maximum subsidence.
These two factors will be taken into account further.

One should also note that the putative melting or thermal erosion
of the terrain during the eruption should not modify the $\alpha_{V}$
value. Indeed, this would only transfer some eroded material from
the underlying terrain to the smooth feature, with a null net weight
balance (as explained in section \ref{subsec:5452r}). For this reason,
the results presented within this study are not affected by putative
melting or thermal erosion of the underlying terrain.

\section{Eruption model}

\subsection{Pressurization by cryomagma freezing \label{subsec:model_basic}}

The model described in this section was presented previously by \citet{Lesage_CryomagmaAscent_2020}.
In this study, we tested the feasibility of a cryomagmatic eruption
model proposed by \citet{Fagents_EuropaCryovolcanism_JGR2003} in
which a cryomagma reservoir is pressurized by its freezing. We modeled
a cryomagma reservoir as a spherical cavity in Europa's ice crust
filled with liquid. Because of the temperature gradient between the
cryomagma and the surrounding ice, cryomagma freezes from the reservoir
wall toward its center, and we model the solidification front position
as a function of time by solving the Stefan problem. The density contrast
between liquid and solid cryomagma generates overpressure in the reservoir,
thus tangential stress on the wall. When the overpressure in the reservoir
is high enough, the wall breaks and a fracture can propagate toward
the surface \citep{Lister_DikeMagmaTransport_1991,Rubin_TensileFractureDikePropagation_1993}.
Then, cryolava can flow onto the surface until the overpressure in
the reservoir has been released.

Based on this model, we can calculate the cryolava volume emitted
at the surface during an eruption and the total duration of this event.
These results are obtained as a function of the cryomagmatic reservoir
parameters (such as its volume $V$ and depth $H$) and its environment
(temperature gradient in the ice crust, ice and cryomagma compositions).
Results and details of this model are available in \citet{Lesage_CryomagmaAscent_2020}.

The model is derived for two cryomagma compositions: 1) pure liquid
water and 2) a briny mixture of 81 wt\% H$_{2}$O + 16 wt\% MgSO$_{4}$
+ 3 wt\% Na$_{2}$SO$_{4}$ that is predicted to be close to the Europa's
ocean and ice composition \citep{Kargel_BrineVolcanism_1991}. This
briny mixture is assumed to be an eutectic composition in the model:
when it freezes, the ice has the same composition and salt content.
The recent detection of chlorides such as NaCl \citep{Trumbo_NaClEuropa_2019}
or Mg-bearing chlorinated species \citep{Ligier_EuropaSurfaceComposition_2016}
on Europa's surface indicates that the actual cryomagma composition
may be a bit different from the one considered here. Nevertheless,
the model results are functions of the density contrast between the
liquid cryomagma and the ice, which is very similar for sulfates and
chlorides \citep{Kargel_BrineVolcanism_1991,Hogenboom_MgSO4_1995,McCarthy_SolidificationHydrates_2007,Quick_HeatTransfertCryomagma_2016,Lesage_CryomagmaAscent_2020}.
So finally, in our model, the salt content is more important than
cryomagma exact composition. Here we take a mixture which contains
19\% of sulfates.

\subsection{Cyclic eruptions \label{subsec:model-cyclic}}

\begin{figure}
\centerline{\includegraphics[width=0.8\textwidth]{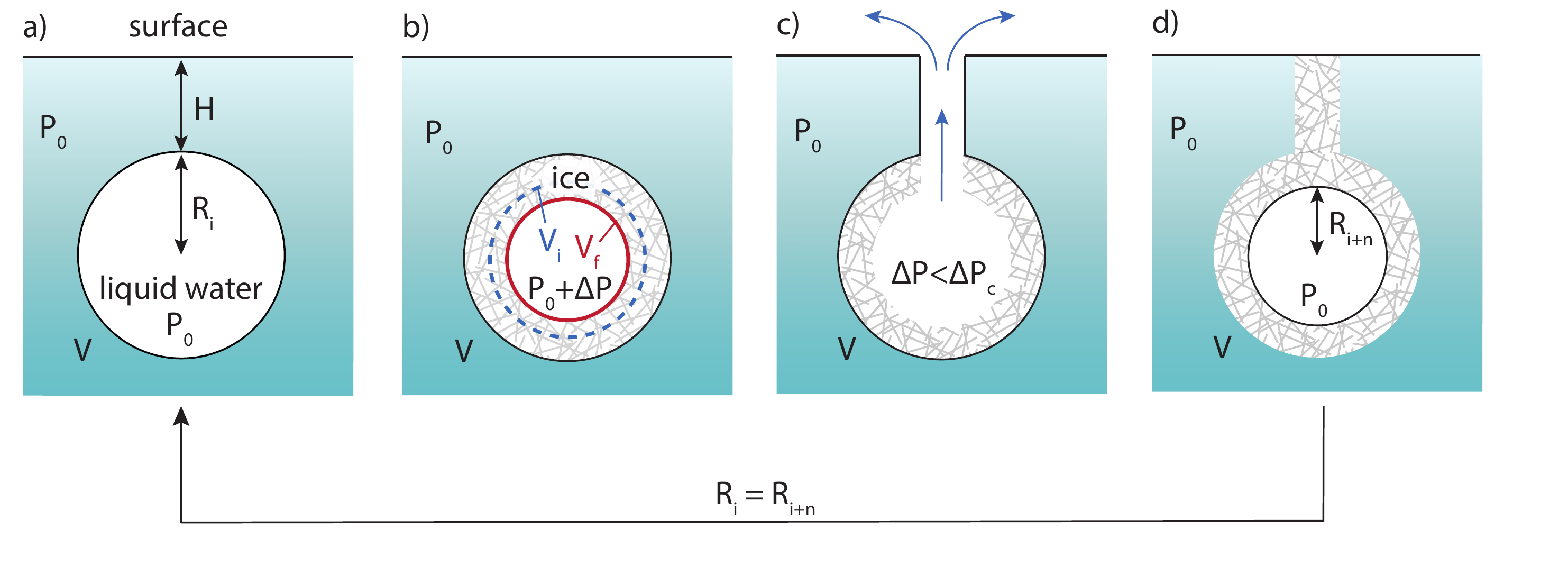}}\caption{Schematic representation of a cryomagma reservoir of volume $V$ and
radius $R=R_{i},$ located at depth $H$ below the surface, with liquid
cryomagma in white and frozen cryomagma in dashed gray. (a) The reservoir
is filled with pure or briny liquid water at isostatic pressure $P_{0}$.
(b) An initial liquid volume $V_{i}$ freezes up to a volume $V_{f}$
of ice, inducing an overpressure $\Delta P$ in the reservoir. (c)
When the pressure reaches a critical value $\Delta P_{c}$, the wall
fractures and the pressurized liquid rises to the surface through
an $H$ long fracture. (d) Once a certain amount of liquid has erupted
at the surface and the overpressure in the reservoir is released,
the eruption ends. The liquid in the fracture freezes, which seals
the reservoir. Finally, the reservoir returns to initial condition
similar to situation (a) but with a smaller radius $R_{i+n}$. Freezing
of the reservoir continues such that these 4 steps repeat and form
an eruptive cycle, leading to several eruptions during the reservoir's
lifetime. \label{fig:model_cyclic} }
\end{figure}

In \citet{Lesage_CryomagmaAscent_2020}, a single cryovolcanic eruption
was modeled. Nevertheless, a reservoir might trigger several eruptions
during its lifetime, as shown in Fig. \ref{fig:model_cyclic}. At
the end of an eruption, the liquid in the fracture may freeze or tectonic
stress may close the reservoir. Nevertheless, the solidification continues
and the freezing front progresses toward the reservoir's center, which
pressurizes the cryomagma once more and can lead to a second eruption.
A reservoir might hence be able to erupt several times. 

\citet{Fagents_EuropaCryovolcanism_JGR2003} predicted that cyclic
eruptive events could result in morphologies where multiple flow lobes
are present. Cyclic eruptions might explain the lobate morphology
that can be seen at the center of the smooth feature from image 9352r
(see Fig. \ref{fig:mosaic_images}d, blue arrow). As this lobate feature
is located on top of the smooth plain, it might have been emplaced
after the smooth plain. We do not clearly see several lobate forms
on the other features, which may be explained by the limited resolution
of the Galileo images. As an example, for image 5452r, the resolution
is approximately 25 m/px. The solar incidence angle is $\sim$75°,
so only features higher than $\sim$11 m can be seen on the image
with their projected shadow, assuming that the shadow is projected
on a flat terrain. Taking into account the complex processes expected
to affect liquid water flowing onto Europa's surface, such as the
possibility of endogenous cryolava flow and the competition between
freezing and vaporizing (see \citealt{Allison_WaterFlowGanymede_1987,Fagents_EuropaCryovolcanism_JGR2003,Quick_CryovolcanicDomes_2016}
and section \ref{subsec:Vaporized-fraction} for details), it is hard
to determine whether the smooth features could result preferentially
from a single or multiple eruptions, so here we consider these two
possibilities.

\begin{figure}
\centerline{\includegraphics[width=1\textwidth]{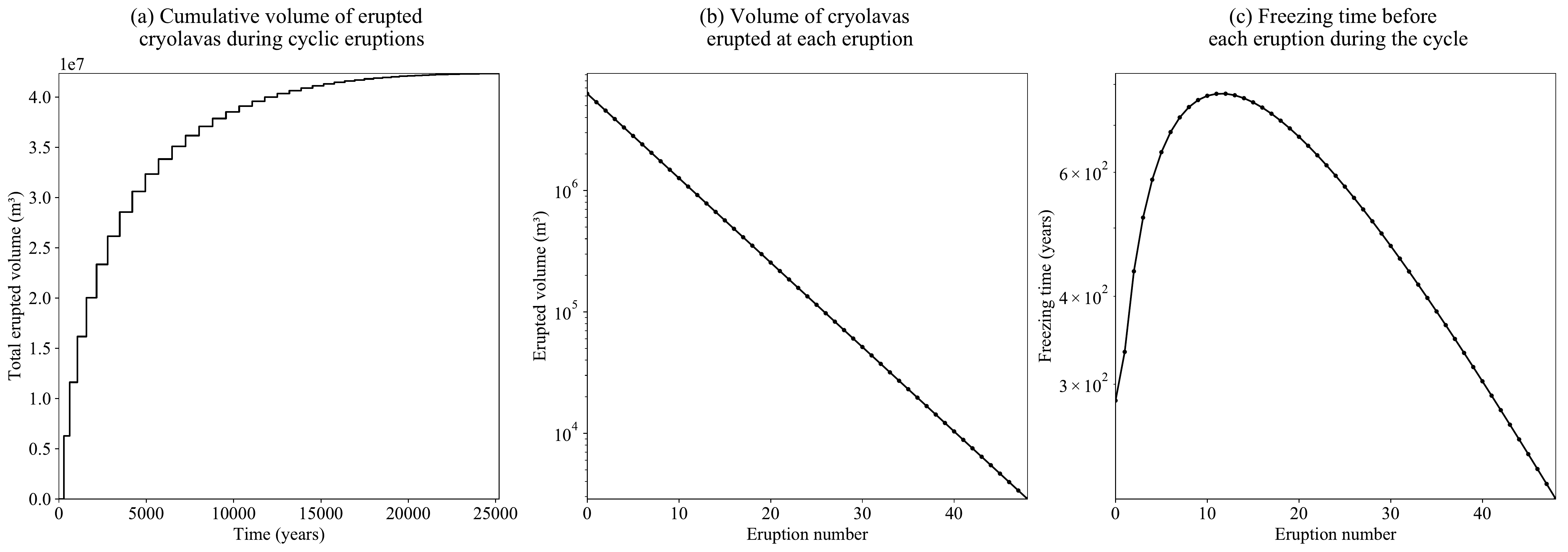}}\caption{(a) Cumulative volume erupted at the surface during the active lifetime
of a reservoir. Each step corresponds to a complete eruptive cycle:
the x-axis represents the freezing time (the eruption time is negligible
compared to the freezing time), and the y-axis stands for the erupted
volume. This result is obtained with a 10$^{9}$ m$^{3}$ reservoir
located 4 km beneath the surface and filled with briny cryomagma.
(b) Volume of cryolava erupted during each eruption as a function
of the eruption number. Note that y-axis is logarithmic here. (c)
Freezing time before each eruption as a function of the eruption number
(this time is non-cumulative).\label{fig:results_1reservoir_cyclical-1}}
\end{figure}

To take into account the possibility of cyclic eruptions, we compare
the smooth feature volumes with the total erupted volume during the
lifetime of a cryomagma reservoir producing cyclic eruptions. To obtain
this volume, we iterate the cryomagmatic eruption model described
in section \ref{subsec:model_basic} as illustrated in \ref{fig:model_cyclic}.
Only the reservoir fraction that remains liquid at the end of each
freezing is taken into account to calculate the erupted cryolava volume.
Fig. \ref{fig:results_1reservoir_cyclical-1}a shows the volume erupted
at the surface during the activity lifetime of a reservoir of 10$^{9}$
m$^{3}$ located 4 km beneath the surface and filled with briny cryomagma.
Each step in Fig. \ref{fig:results_1reservoir_cyclical-1}a represents
an eruptive cycle, i.e. cryomagma freezing, pressurization, and eruption.

Fig. \ref{fig:results_1reservoir_cyclical-1}b shows the volume of
cryolava erupted at the surface during each eruption. It follows a
decreasing logarithmic trend (note that y-axis is logarithmic). We
can predict the volume $V_{\#i}$ erupted during each eruption as
following. For the first eruption, the erupted volume is equal to
the increase in the frozen cryomagma volume. The freezing cryomagma
volume is equal to $nV$ where $n$ is the cryomagma fraction necessary
to freeze in order to trigger an eruption \citep{Fagents_EuropaCryovolcanism_JGR2003,Lesage_CryomagmaAscent_2020}
and $V$ is the reservoir initial volume. The cryomagma volume that
does not freeze is defined as $V_{0i}=V(1-n)$. The newly formed ice
has a volume $n\frac{\rho_{l}}{\rho_{s}}V$ where $\rho_{l}$ is the
cryomagma density and $\rho_{s}$ is the ice density, so the liquid
part of the reservoir is compressed to a volume $V_{0f}=V(1-n\frac{\rho_{l}}{\rho_{s}})$
(see \citealp{Lesage_CryomagmaAscent_2020} for details). Finally
the volume change of the liquid part in the reservoir after freezing
is \citep{Fagents_EuropaCryovolcanism_JGR2003,Lesage_CryomagmaAscent_2020}:
\begin{equation}
\begin{alignedat}{1}V_{\#0} & =V_{0i}-V_{0f}\\
 & =nV\left(\frac{\rho_{l}}{\rho_{s}}-1\right)
\end{alignedat}
\label{eq:volume_increase_freezing}
\end{equation}
Then, for the second eruption, we calculate $V_{1i}$ and $V_{1f}$
from the remaining cryomagma after the first eruption, which has a
volume $V_{1i}=V_{0f}(1-n)$. The erupted volume $V_{\#1}$ is:
\begin{equation}
\begin{alignedat}{1}V_{\#1} & =V_{1i}-V_{1f}\\
 & =V_{0f}\left(1-n\right)-V_{0f}\left(1-n\frac{\rho_{l}}{\rho_{s}}\right)\\
 & =V_{\#0}\left(1-n\frac{\rho_{l}}{\rho_{s}}\right)
\end{alignedat}
\label{eq:V1}
\end{equation}
Eq. (\ref{eq:V1}) can be generalized for all the following eruptions,
so finally the erupted volume at eruption $\#i$ can be written:

\begin{equation}
V_{\#i}=V_{\#0}\left(1-n\frac{\rho_{l}}{\rho_{s}}\right)^{\#i}\label{eq:Vi_f(V0)}
\end{equation}
 or:

\begin{equation}
V_{\#i}=nV\left(\frac{\rho_{l}}{\rho_{s}}-1\right)\left(1-n\frac{\rho_{l}}{\rho_{s}}\right)^{\#i}\label{eq:Vi}
\end{equation}

Fig. \ref{fig:results_1reservoir_cyclical-1}c demonstrates that the
time between each eruption initially increases before decreasing.
In fact, this freezing time is a competition between two phenomena:
firstly, the thermal transfer between the warm reservoir and the cold
surrounding ice slows down over time as the reservoir gradually cools,
so the thermal wave propagates slower; secondly, the liquid volume
decreases, which reduces the amount of cryomagma that needs to freeze
to trigger an eruption. For the example given in Fig \ref{fig:results_1reservoir_cyclical-1}c,
the first phenomenon dominates from eruptions \#1 to \#12, then the
second one becomes preponderant. To explain this trend, one must compare
two characteristic velocities: the first one is the solidification
front progression, and the second one is the heat transfer in the
ice. These two velocities are obtained by solving the Stefan problem
at the reservoir wall and thus are different for each eruption.

\begin{figure}
\centerline{\includegraphics[width=1\textwidth]{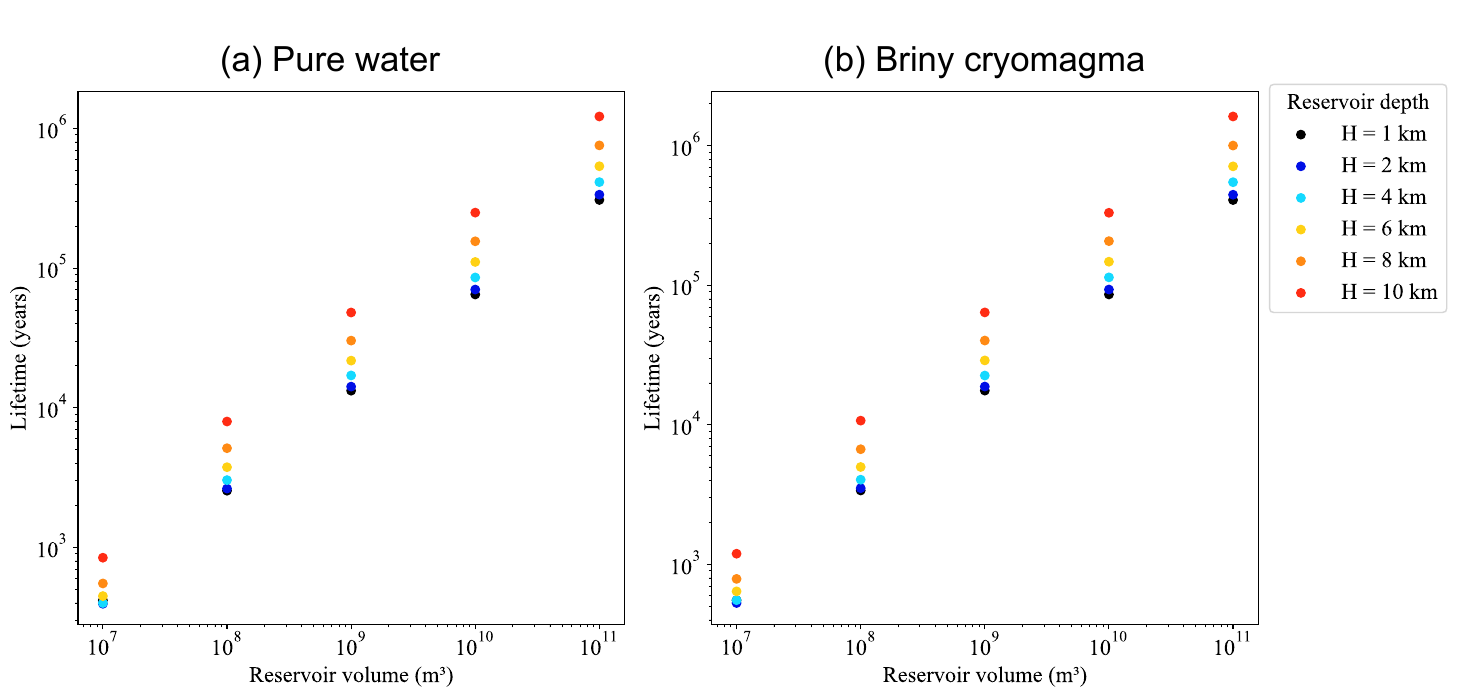}}\caption{Total activity time of a cryomagmatic reservoir producing cyclic eruptions
as a function of the reservoir volume and depth. Two cryomagma compositions
are tested: (a) pure water and (b) a briny mixture of 81 wt\% H$_{2}$O
+ 16 wt\% MgSO$_{4}$ + 3 wt\% Na$_{2}$SO$_{4}$. The reservoir is
active until all the cryomagma is frozen.\label{fig:results_model_cyclical_time}}
\end{figure}

Fig. \ref{fig:results_model_cyclical_time} shows the total activity
lifetime of a reservoir, considering that one cryomagmatic reservoir
may erupt several times. The reservoir activity lifetime is calculated
for 5 different reservoir volumes ranging from $10^{6}$ to $10^{11}$
m$^{3}$ and for 6 different depths ranging from $1$ to $10$ km
under the surface. Lifetime increases with the reservoir volume, for
both pure and briny cryomagma, ranging from 0.4 years to $10^{5}$
years. In addition, reservoir lifetime is approximately 10 times larger
for reservoir at 10 km depth in comparison to 1 km depth. Reservoir
lifetime depends on reservoir depth because of the temperature gradient
in the ice crust: reservoirs near the surface are located in a colder
environment, which makes them freeze faster.

\begin{figure}
\centerline{\includegraphics[width=0.5\textwidth]{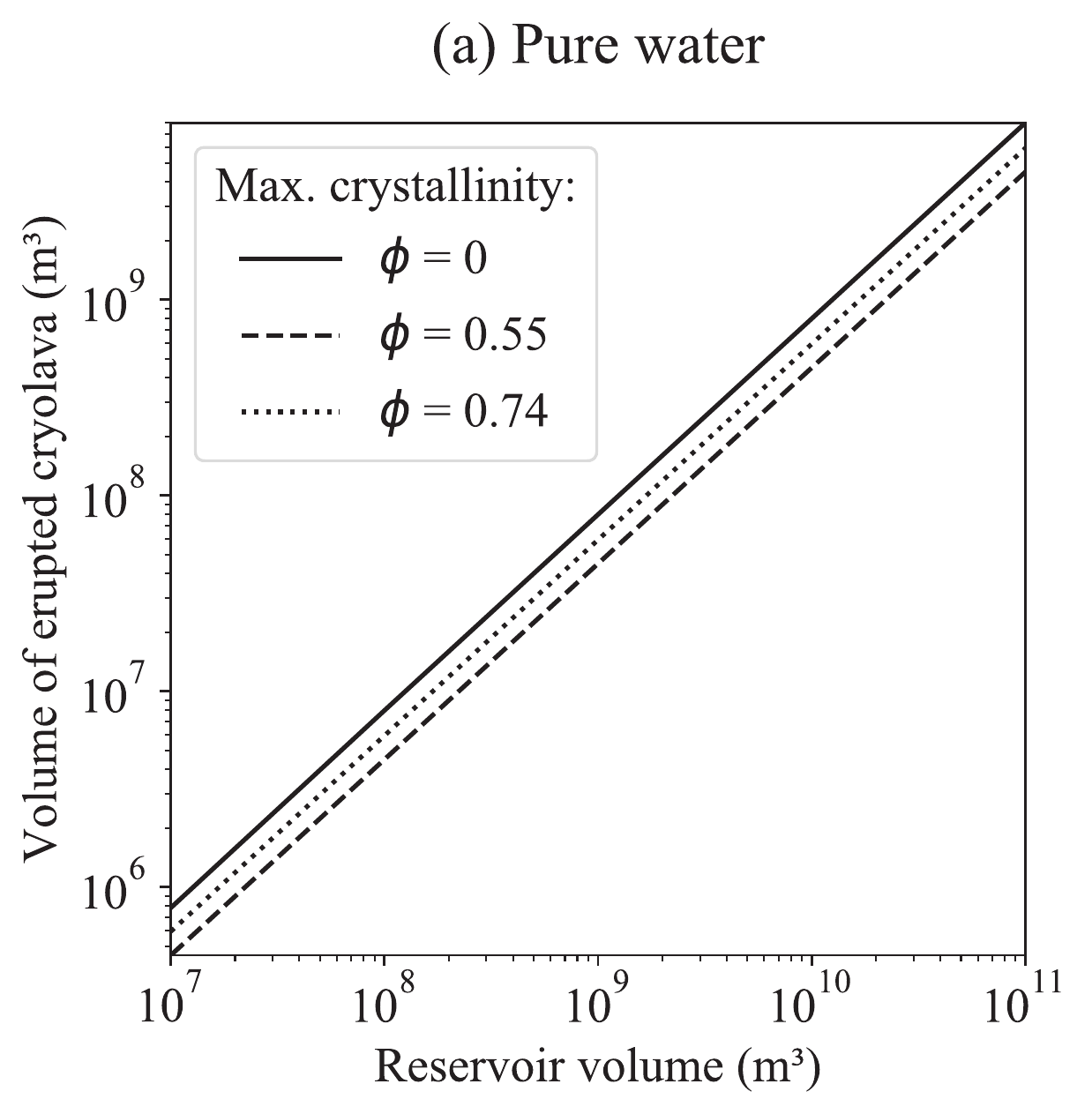}\includegraphics[width=0.5\textwidth]{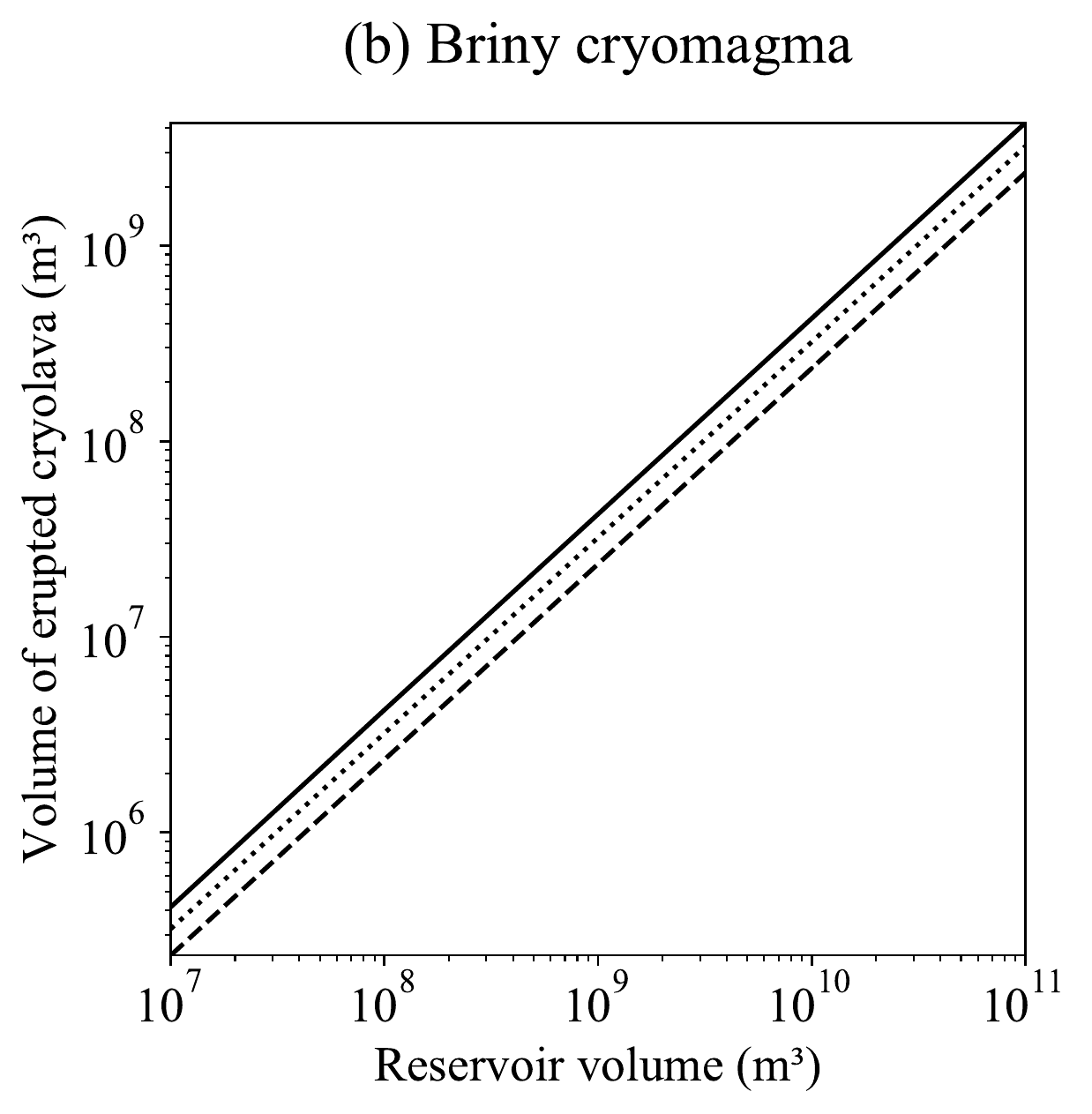}}\caption{Total cryolava volume erupted from a reservoir producing cyclic eruptions
as a function of reservoir volume. Two cryomagma compositions are
used: (a) pure water and (b) a briny mixture of 81 wt\% H$_{2}$O
+ 16 wt\% MgSO$_{4}$ + 3 wt\% Na$_{2}$SO$_{4}$. The reservoir is
active until the volume of ice in it reaches the cryomagma crystallinity
$\phi$ times the reservoir volume $V$ (or simply the reservoir volume
$V$ in the case of $\phi$ =0).\label{fig:results_model_cyclical_volumes}}
\end{figure}

In the model presented above, we considered that a cryomagma reservoir
may remain active as long as some liquid remains in its interior.
Nevertheless, as cryomagma freezes, the crystal concentration in the
liquid might increase, which increases the effective cryomagma viscosity
\citep{Roscoe_Viscosity_1952}. The Einstein--Roscoe law \citep{Roscoe_Viscosity_1952}
gives the effective viscosity $\mu_{eff}$ of a fluid and crystals
mixture as a function of the fluid viscosity $\mu$ and the crystal
concentration $\phi$ assuming spherical crystals:
\begin{equation}
\mu_{eff}=\mu\left(1-1.35\phi\right)^{-2.5}\label{eq:Roscoe_formula}
\end{equation}
 \citet{Marsh_Cristallynity_1981} studied the effect of crystals
in terrestrial magmas, and based on the Einstein--Roscoe law, calculated
that eruption of lava should stop once 50 to 60\% of crystallinity
is reached due to a large increase in viscosity. \citet{Quick_CryovolcanicDomes_2016}
proposed to apply this criterion to cryomagma on Europa and consider
that cryomagma eruption is not possible for crystal content above
$\sim$55\%. Nevertheless, cryomagma on Europa is expected to be a
water-based mixture \citep{Kargel_BrineVolcanism_1991}, with a viscosity
lower than terrestrial magmas of several orders of magnitude. To calculate
the maximum crystallinity allowing cryomagma to erupt, we use Eq.
(\ref{eq:Roscoe_formula}) and effective viscosity measurements of
cryovolcanically emplaced domes done by \citet{Quick_CryovolcanicDomes_2016}.
They deduced from those measurements that cryolava should have an
effective viscosity $\mu_{eff}$ ranging from 1 to 10$^{4}$ Pa.s
during the eruption to create dome features. Based on this result,
and using the liquid water viscosity $\mu\simeq10^{-3}$ Pa.s as the
cryolava liquid phase viscosity, we can deduce an upper limit of crystallinity
of $\phi\simeq0.74$ for erupted cryomagma, assuming that the crystals
were the only effect acting to increase the cryolava viscosity at
the surface.

As eruptions are expected to stop when cryomagma reaches a maximum
crystallinity, this threshold can be considered as a stop condition
for our cyclic eruption model. Fig. \ref{fig:results_model_cyclical_volumes}
shows the total volume of cryolava erupted during the reservoir lifetime
for three different maximum cryolava crystallinity: $\phi$=0, 0.55
and 0.74. $\phi$=0 corresponds to cryolava that does not contain
crystals. This could occur if crystals are totally separated from
cryomagma, if they remain on the reservoir wall for example. In this
case, crystals do not ascend with the fluid and cryolava keeps a very
low viscosity, close from the pure water viscosity. Because of the
very smooth and thin appearance of the features studied here, $\phi$
could potentially be close to 0. To model this extreme case, we continue
the reservoir freezing until all the cryomagma is turned to ice. $\phi$=0.55
is the maximal cryolava crystallinity suggested by \citet{Quick_CryovolcanicDomes_2016}
according to the study of \citet{Marsh_Cristallynity_1981}. To model
this maximum crystal concentration in cryolava, we stop the reservoir
activity when 55\% of reservoir volume is solid. Finally, $\phi$=0.74
corresponds to the crystallinity calculated from Eq. (\ref{eq:Roscoe_formula})
and the maximum cryolava viscosity obtained by \citet{Quick_CryovolcanicDomes_2016}.
We model this threshold by stopping the reservoir activity when 74\%
of reservoir volume is solid. For reservoir volumes ranging from 10$^{7}$
to 10$^{11}$, we finally obtain erupted volumes ranging from 5$\times$10$^{5}$
to 7$\times$10$^{9}$ m$^{3}$ for pure water (Fig. (\ref{fig:results_model_cyclical_volumes})a),
and from 2$\times$10$^{5}$ to 3$\times$10$^{9}$ m$^{3}$ for the
briny cryomagma. To obtain the following results, we use $\phi$=0
as it seems to be in better agreement with the very thin topography
of smooth plains studied here. Nevertheless, one should keep in mind
that the erupted volumes may be divided by at most a factor two if
a more viscous cryomagma is involved.

\subsection{Vaporized fraction of water \label{subsec:Vaporized-fraction}}

The putative flow of water-based liquid on Europa takes place in a
low pressure and temperature environment. The pressure at Europa's
surface is near 10$^{-6}$ Pa \citep{Hall_OxygenAtmEuropa_1995} and
the mean temperature is around 110 K \citep{Spencer_EuropaSurfaceTemperature_1999},
so that the liquid water erupting at the surface is subjected to the
counteractions of freezing and boiling. Because of the $\simeq160$
K difference between the liquid and the environment, several authors
previously proposed that an ice crust forms rapidly on top of the
flow \citep{Allison_WaterFlowGanymede_1987,Fagents_EuropaCryovolcanism_JGR2003,Quick_CryovolcanicDomes_2016}.
\citet{Allison_WaterFlowGanymede_1987} studied the flow of liquid
water on Ganymede's surface and found that a $\sim0.5$ m thick ice
crust is enough to prevent the flow from boiling, which allows the
underlying liquid to flow onto the surface. \citet{Quick_CryovolcanicDomes_2016}
calculated crust thickness as a function of the time for erupted brines
on Europa and found that a 0.5 m thick crust would form in $\sim$7.5
days. Before the formation of a solid crust on top of the flow, the
liquid would boil violently. Moreover, once the crust is formed, it
would stabilize the interior portion of the flow \citep{Allison_WaterFlowGanymede_1987,Quick_CryovolcanicDomes_2016},
but the expanding edges are supposed to look like a mix of boiling
water and ice blocks being pushed by the liquid \citep{Allison_WaterFlowGanymede_1987}.
It is thus necessary to estimate the vaporized fraction of cryolava
during the eruption.

The water fraction being vaporized at the surface during the cryolava
flow is a key parameter to link the volume of the smooth features
and the initial erupted volume of liquid water, hence, it must be
evaluated. This quantity was roughly estimated by \citet{Porco_CassiniEnceladus_2006},
who calculated the vaporized fraction of water during the opening
of cracks on Enceladus. One should note that this study does not take
into account the formation of an ice crust as discussed above, so
the amount of material vaporized used in this study can be considered
as an upper boundary. \citet{Porco_CassiniEnceladus_2006} assumes
that the latent heat of fusion $L_{f}$ generated by the freezing
part of liquid water is used as latent heat of vaporization $L_{v}$
by the vaporized part of the fluid, so they calculate that a fraction
$x=L_{f}/(L_{f}+L_{v})=0.13$ of liquid is vaporized. 

\begin{figure}
\centerline{\includegraphics[width=0.8\textwidth]{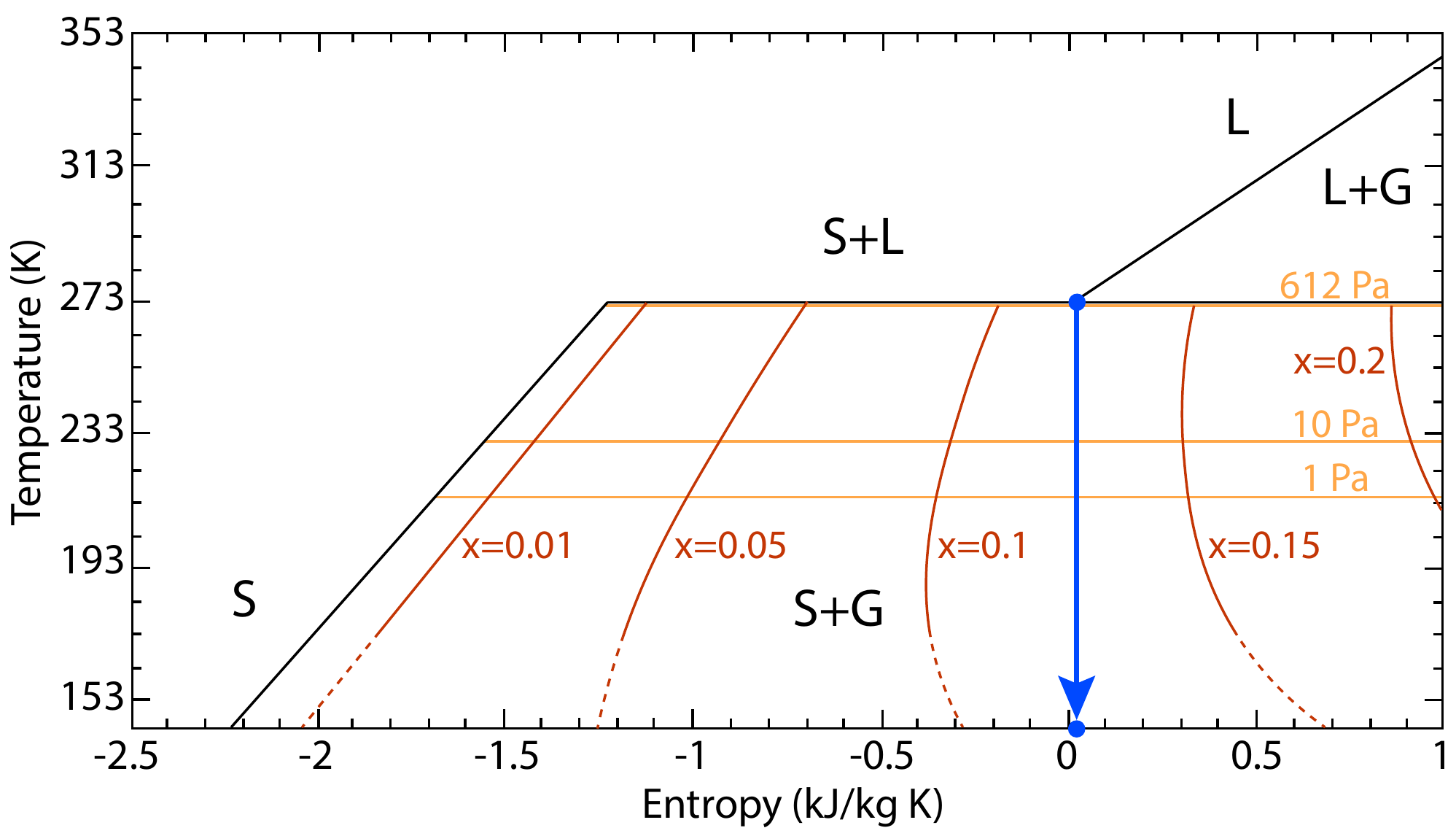}}\caption{Temperature--entropy diagram of pure water. S stands for ``solid'',
L for ``liquid'' and G for ``gas''. $x$ is the vapor ratio in
the solid + gas mixture. The blue arrow shows the path followed by
liquid water at 273 K coming from a sub-surface reservoir (T=273 K,
P=612 Pa the pressure at the triple point) and flowing on the surface
of an icy moon (low pressure and temperature). This process takes
place between the $x$=0.1 and $x$=0.15 vapor ratio lines. This diagram
is adapted from \citet{Lu_ThermoH2O_2009}. \label{fig:diagram_E-T}}
\end{figure}

This simple calculation from \citet{Porco_CassiniEnceladus_2006}
gives an idea of the quantity of liquid turned into vapor when it
reaches the icy moon's surface, but to have a better knowledge of
the transformation occurring in the two-phase region it is necessary
to use the phase diagram of water. To assess the water behavior in
extreme environments such as Europa's surface, it is relevant to use
a temperature--entropy (T-s) phase diagram \citep{Lu_ThermoH2O_2009}
as the flow might be considered as isentropic (adiabatic and reversible)
as discussed by \citet{Kieffer_IsentropicDecompressionFluids_1979}.
An isentropic process is a vertical line on a T-s phase diagram, so
it is easy to deduce flow properties based on this representation,
and one can directly read the mass ratio of gas/solid on such a diagram.
Fig. \ref{fig:diagram_E-T} shows the T-s diagram of water adapted
from \citet{Lu_ThermoH2O_2009}. On this diagram, the blue arrow is
the isentropic depressurizing process from liquid water at the triple
point (T=273 K, P=612 Pa) to the exit state, i.e. the conditions at
Europa's surface (low pressure and low temperature). This process
takes place between the $x$=0.1 and $x$=0.15 vapor fraction lines
on Fig. \ref{fig:diagram_E-T}. It indicates that at the end of the
process, only 10 to 15\% of the liquid water erupted at the surface
is vaporized; the major fraction of water freezes to solid. 

These results show that, without taking into account the formation
of an ice crust on top of the flow, the vaporized fraction of water
should range between 10 and 15\% of the total erupted cryolava. The
ice crust formed on top the of the flow plays a role in protecting
the well-developed flow from boiling, however, the flow edges are
still subject to boiling and freezing counteractions. We thus consider
a vaporized fraction of 13$\pm$3\% after \citet{Porco_CassiniEnceladus_2006}
and \citet{Lu_ThermoH2O_2009} to calculate the liquid volume that
may be at the origin of the smooth features, keeping in mind that
this calculated volume is a lower limit of the erupted volume. Also,
the addition of salts or impurities in the cryomagma could slightly
modify this result as it lowers the vapor pressure on Europa from
$\sim$600 to 300-500 Pa depending on the salt content \citep{Quick_CryovolcanicDomes_2016}.

Another parameter that could affect the density contrast between the
liquid and solid cryomagma phases is the formation of hydrates during
the solidification process. In our previous work \citep{Lesage_CryomagmaAscent_2020},
we considered a briny cryomagma composed of the following mixture:
81 wt\% H$_{2}$O + 16 wt\% MgSO$_{4}$ + 3 wt\% Na$_{2}$SO$_{4}$,
which is the composition of Europa's ocean and ice predicted by \citet{Kargel_BrineVolcanism_1991}.
\citet{McCarthy_SolidificationHydrates_2007} show that hydrates of
MgSO$_{4}$ and Na$_{2}$SO$_{4}$ form for concentrations above respectively
17.3 and 4 wt\%. Thus, in the cases considered in this work, hydrates
should not form in the freezing cryomagma. Nevertheless, hydrates
formation must be taken into in the case of a higher salt concentration.

\section{Results}

\subsection{Measured and erupted volumes \label{subsec:Measured_and_erupted_volumes}}

Table \ref{tab:results} summarizes smooth feature volumes measured
on the four images shown in Fig. 1. Measured volumes $V_{measured}$
are extracted directly from the DEM (see the method in section \ref{subsec:Gis}
and flowchart in Fig. \ref{fig:flowchart_method}). As detailed previously,
the measured volume is not equal to the volume of cryomagma erupted
at the surface ($V_{erupted}$ ) during the eruption that created
the feature. To obtain the erupted volume, we multiply the measured
volume by the $\alpha_{V}$ factor to take into account the ridges
on the covered surface and putative subsidence (see section \ref{subsec:alpha}
and Fig. \ref{fig:topo_ridges}). We calculate the results for two
extreme values of $\alpha_{V}$: $\alpha_{V}=0.8$, which is the case
of shallow subsidence, and $\alpha_{V}=5$, which describes the maximum
subsidence possibly induced by a liquid subsurface reservoir of 5
km radius according to \citet{Manga_Lenticulae_2016}. We also multiply
the volume $V_{measured}$ by a factor $\nicefrac{1}{1-x}$ (where
$x$ is the cryomagma vaporized fraction) to take into account the
vaporization of the erupted liquid. Finally, we multiply $V_{measured}$
by a factor $\frac{\rho_{s}}{\rho_{l}}$ ($\frac{1130}{1180}$ for
briny cryomagma or $\frac{920}{1000}$ for pure water) to estimate
the liquid volume before expansion due to phase change. This process
is summarized in the following equation: 
\begin{equation}
V_{erupted}=\alpha_{V}V_{measured}\frac{1}{(1-x)}\frac{\rho_{s}}{\rho_{l}}\label{eq:V_erupt}
\end{equation}
where $V_{erupted}$ is the fluid volume erupted from the reservoir,
$V_{measured}$ is the smooth feature volume measured from the DEM
using the simple approach, $\alpha_{V}$ is a volume factor to take
into account the underlying terrain (see section \ref{subsec:alpha}),
$x=0.13$ is the fluid vaporized fraction, $\rho_{l}$ is the density
of the liquid cryomagma and $\rho_{s}$ is the density of the corresponding
ice.

We propagate the uncertainties from the DEM in order to take into
account the two main uncertainty sources, i.e. the smoothness coefficient
and the reflectance model (and the associated albedo). Uncertainty
calculation using mean deviation are detailed in supplementary materials,
section 2. We found an uncertainty of $\pm$15\% on the volumes measured
from the DEM. Moreover, we added a $\pm$3\% uncertainty on the calculation
of the erupted volume due to the uncertainty on the vaporized fraction
(see section \ref{subsec:Vaporized-fraction}). The DEM of each smooth
feature used to calculate the volumes are showed in supplementary
materials, section 3. Finally, we obtain the results given in Table
\ref{tab:results}.

\begin{table*}
\centerline{%
\begin{tabular*}{1\textwidth}{@{\extracolsep{\fill}}c|>{\raggedright}m{3cm}>{\centering}m{4cm}>{\centering}m{3cm}}
\hline 
\noalign{\vskip0.2cm}
\multirow{2}{*}{Image} & \multirow{2}{3cm}{Measured volume (m$^{3}$)} & \multicolumn{2}{>{\centering}p{8cm}}{Erupted volume for pure water (m$^{3}$) (see Eq. \ref{eq:V_erupt}) }\tabularnewline
\cline{3-4} 
 &  & $\alpha_{V}=0.8$ & $\alpha_{V}=5$\tabularnewline
\hline 
5452r & (5.7$\pm$0.9)$\times$10$^{7}$ &  (4.9$\pm$0.9)$\times$10$^{7}$ & (3.0$\pm$0.5)$\times$10$^{8}$\tabularnewline
0713r & (6.6$\pm$1.0)$\times$10$^{7}$ &  (5.6$\pm$1)$\times$10$^{7}$ & (3.5$\pm$0.6)$\times$10$^{8}$\tabularnewline
9352r & (4.2$\pm$0.6)$\times$10$^{8}$ &  (3.5$\pm$0.6)$\times$10$^{8}$ & (2.2$\pm$0.4)$\times$10$^{9}$\tabularnewline
0739r & (2.7$\pm$0.4)$\times$10$^{8}$ &  (2.3$\pm$0.4)$\times$10$^{8}$ & (1.4$\pm$0.3)$\times$10$^{9}$\tabularnewline
\hline 
\end{tabular*}}

\centerline{%
\begin{tabular*}{1\textwidth}{@{\extracolsep{\fill}}c|>{\raggedright}m{3cm}>{\centering}m{4cm}>{\centering}m{3cm}}
\hline 
\noalign{\vskip0.2cm}
\multirow{2}{*}{Image} & \multirow{2}{3cm}{Measured volume (m$^{3}$)} & \multicolumn{2}{>{\centering}p{8cm}}{Erupted volume for briny cryomagma (m$^{3}$) (see Eq. \ref{eq:V_erupt}) }\tabularnewline
\cline{3-4} 
 &  & $\alpha_{V}=0.8$ & $\alpha_{V}=5$\tabularnewline
\hline 
5452r & (5.7$\pm$0.9)$\times$10$^{7}$ &  (5.0$\pm$0.9)$\times$10$^{7}$ & (3.2$\pm$0.5)$\times$10$^{8}$\tabularnewline
0713r & (6.6$\pm$1)$\times$10$^{7}$ &  (5.8$\pm$1)$\times$10$^{7}$ & (3.7$\pm$0.6)$\times$10$^{8}$\tabularnewline
9352r & (4.2$\pm$0.6)$\times$10$^{8}$ &  (3.7$\pm$0.7)$\times$10$^{8}$ & (2.3$\pm$0.4)$\times$10$^{9}$\tabularnewline
0739r & (2.7$\pm$0.4)$\times$10$^{8}$ &  (2.4$\pm$0.4)$\times$10$^{8}$ & (1.5$\pm$0.3)$\times$10$^{9}$\tabularnewline
\hline 
\end{tabular*}}\caption{Summary of the measured volumes $V_{measured}$ using the DEMs, and
the corresponding erupted volumes $V_{erupted}$ calculated with Eq.
(\ref{eq:V_erupt}) for 5 m subsidence ($\alpha_{V}=0.8$) and 40
m subsidence ($\alpha_{V}=5$).\label{tab:results}}
\end{table*}

\subsection{Volume of cryomagmatic reservoirs }

We previously obtained cryolava volume necessary to explain the emplacement
of the four smooth features from Fig. \ref{fig:mosaic_images}. We
can now deduce the reservoir volume required to generate this cryolava
amount from our eruption model.

\begin{figure}
\centerline{\includegraphics[width=1\textwidth]{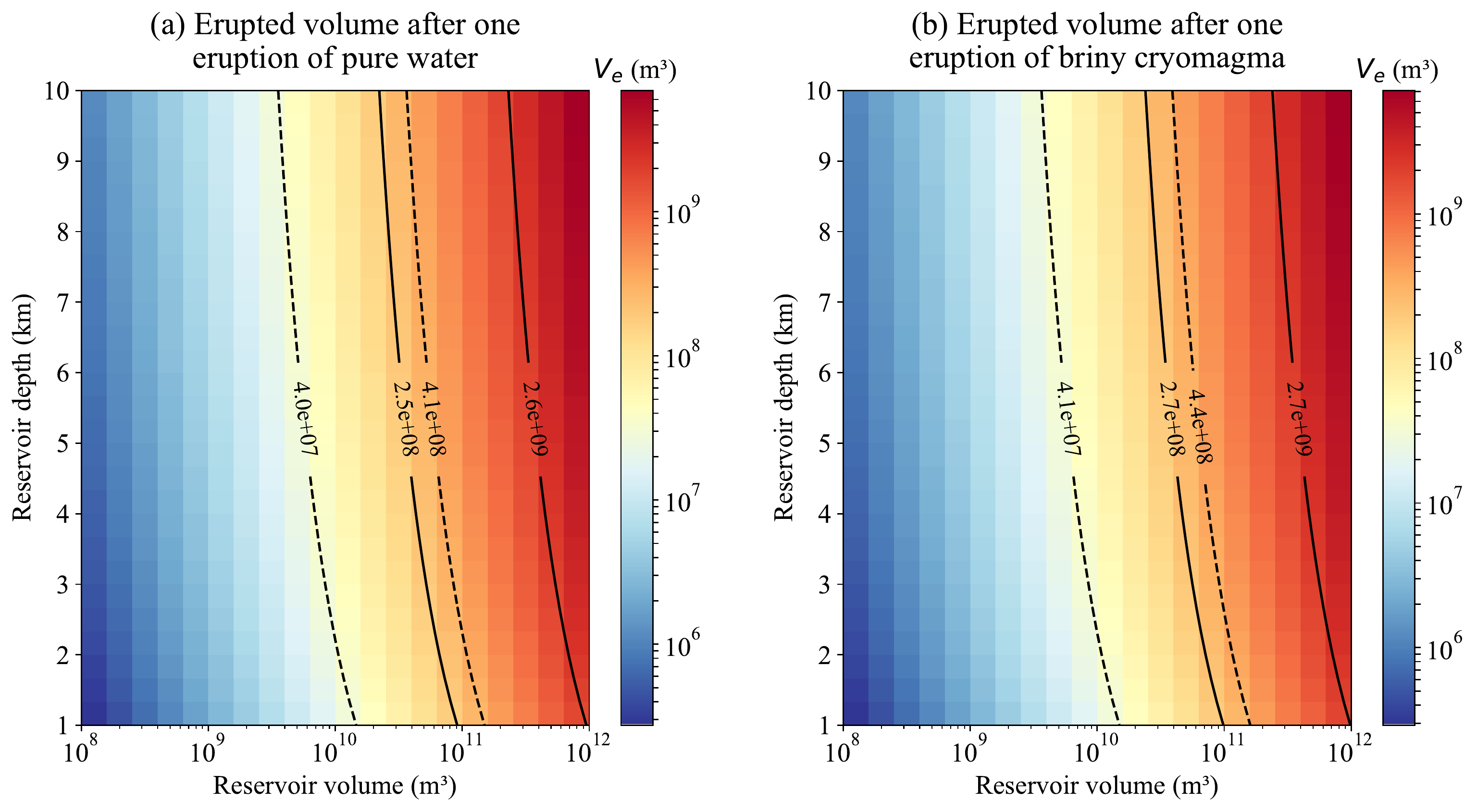}}\caption{Volume erupted at the surface during one cryovolcanic eruption after
the model from \citet{Lesage_CryomagmaAscent_2020}. The smooth feature
volume measured on the four analyzed images (see Fig. \ref{fig:mosaic_images})
corresponds to reservoir volumes ranging in between the two dashed
lines for $\alpha_{V}=0.8$ and between the two solid lines for $\alpha_{V}=5$
(see Table 1). \label{fig:results_model_simple} }
\end{figure}

In \citet{Lesage_CryomagmaAscent_2020}, we obtained the volume of
cryolava erupted at the end of a single eruptive event as a function
of reservoir volume and depth, for two different cryomagma compositions:
pure water and a briny mixture of water and salts: 81 wt\% H$_{2}$O
+ 16 wt\% MgSO$_{4}$ + 3 wt\% Na$_{2}$SO$_{4}$ \citep{Kargel_BrineVolcanism_1991}.
These volumes are shown in Fig. \ref{fig:results_model_simple} for
a reservoir depth ranging from 1 to 10 km and a reservoir volume ranging
from 10$^{8}$ to 10$^{12}$ m$^{3}$, which corresponds to reservoir
radius between $\sim$0.3 and 6.2 km. In Fig. \ref{fig:results_model_simple},
we compare these results with the erupted volumes measured from the
four DEMs (uncertainties on the erupted volumes are taken into account).
One can see in Fig. \ref{fig:results_model_simple} that the eruption
of a 3$\times$10$^{9}$ to 10$^{11}$ m$^{3}$ reservoir (0.9 to
2.9 km radius) is necessary to explain the formation of smooth features
from a single cryovolcanic event for shallow terrain subsidence ($\alpha_{V}=0.8$).
For deeper subsidence ($\alpha_{V}=5$), a 2$\times$10$^{10}$ to
10$^{12}$ m$^{3}$ reservoir (1.7 to 6.2 km radius) is required.
The range of erupted volumes is nearly identical for these two compositions
as the $\frac{\rho_{ice}}{\rho_{liq}}$ factor does not differ significantly
between pure water and a water-based mixture. The erupted volume necessary
to produce the observed features depends mostly on the $\alpha_{V}$
ratio. One should note that larger reservoirs are expected to create
deeper subsidence of the surface \citep{Manga_Lenticulae_2016}, so
it is worth it to consider both solutions.

In addition, we also consider in this study the case of several eruptive
cycles instead of one single eruption (see section \ref{subsec:model-cyclic}).
In this case, the same reservoir will produce several eruptions during
its whole lifetime. According to Fig. \ref{fig:results_model_cyclical_time},
this kind of reservoir may erupt 5$\times$10$^{6}$ to 7$\times$10$^{9}$
m$^{3}$ of cryolava if we consider that the reservoir is active until
100\% of the cryomagma is frozen. One should note that this amount
may be overestimated as eruption may stop when 55\% of the reservoir
is frozen \citep{Quick_CryovolcanicDomes_2016}. This could divide
cryolava erupted volume by a factor $\sim$2, but is not taken into
account in the following results as cryolava without crystal is in
better agreement with the smooth feature morphologies. We summarize
in Table \ref{tab:results_cyclic} the reservoir volumes required
to obtain the minimum and maximum erupted volumes given in Table \ref{tab:results}
(respectively for image 5452r and 9352r), taking uncertainties into
account. 
\begin{table}
\hfill{}%
\begin{tabular}{|c|>{\centering}p{3cm}|>{\centering}p{2.5cm}|>{\centering}p{2.5cm}|}
\cline{3-4} 
\multicolumn{1}{c}{} &  & Pure water & Briny cryomagma\tabularnewline
\hline 
\multirow{2}{*}{$\alpha_{V}=0.8$} & Min. reservoir size (image 5452r) & 4$\times$10$^{8}$

($R$ = 0.5 km) & 10$^{9}$

($R$ = 0.6 km)\tabularnewline
\cline{2-4} 
 & Max. reservoir size (image 9352r) & 4 $\times$10$^{9}$

($R$ = 1 km) & 10$^{10}$

($R$ = 1.3 km)\tabularnewline
\hline 
\multirow{2}{*}{$\alpha_{V}=5$} & Min. reservoir size (image 5452r) & 2$\times$10$^{9}$

($R$ = 0.8 km) & 6$\times$10$^{9}$

($R$ = 1.2 km)\tabularnewline
\cline{2-4} 
 & Max. reservoir size (image 9352r) & 2$\times$10$^{10}$

($R$ = 1.6 km) & 6 $\times$10$^{10}$

($R$ = 2.4 km)\tabularnewline
\hline 
\end{tabular}\hfill{} \caption{Size of the reservoir required to erupt cryolava amount necessary
to generate smooth features observed depending on cryomagma composition
and $\alpha_{V}$ ratio. We give these results for the smallest and
largest features, respectively from image 5452r and 9352r. We use
the following composition for the briny cryomagma: 81 wt\% H$_{2}$O
+ 16 wt\% MgSO$_{4}$ + 3 wt\% Na$_{2}$SO$_{4}$ \citep{Kargel_BrineVolcanism_1991}.
\label{tab:results_cyclic}}
\end{table}

\section{Discussion and conclusions \label{sec:Discussion}}

We identified four images from Galileo SSI data presenting smooth
features that may have formed during one or several eruptions of cryolava
at Europa's surface. We produced DEMs of those smooth features using
the Shape from Shading tool of the AMES Stereo Pipeline, and we measured
their volumes and associated uncertainties. The major uncertainties
on the DEM generation and volume measurement come from the user parameters
of SfS, but their effect is relatively low for small scale features.
We estimated the uncertainty on the volumes at around $\pm$15\% (see
supplementary materials, section 2). Volume measurements of the four
selected smooth features gave results ranging from 5$\times$10$^{7}$
to 5$\times$10$^{8}$ m$^{3}$.

Shape from Shading tool allowed us to generate DEMs from single images,
with high precision at low scales \citep{Nimmo_StereoSfSCompare_2008}.
Nevertheless, SfS presents its own limitations. First of all, it does
not manage albedo or photometric heterogeneities in a single image.
In fact, the reflectance model chosen by the user is applied to the
whole image, but the surface properties can differ at regional or
local scales, as shown by \citet{Jiang_ASPSfSandStereo_2017} and
\citet{Belgacem_RegionalPhotometryEuropa_2019}. We showed that the
reflectance model and the associated chosen coefficients (such as
albedo) are not expected to create uncertainties on volume higher
than $\pm$10\% for small scale features. Nevertheless, as discussed
by \citet{Jiang_ASPSfSandStereo_2017} for Martian images, too many
heterogeneous surface properties in one image can lead to non-convergence
of the algorithm. From the images we selected, four were converging
(images presented in Fig. \ref{fig:mosaic_images}), but it was not
the case of image 8613r, which contains a smooth material darker than
the surrounding terrain, maybe due to an albedo heterogeneity. We
could not provide a DEM for this image. It could be interesting to
investigate the possibility of considering several zones with different
surface properties in an image: this could solve the problem of DEM
generation of image 8613r and give information on this smooth feature.

Geomorphological interpretations of DEMs are consistent with smooth
feature formation by the flow of a fluid on the surface: smooth features
have a very thin appearance, are constrained by the surrounding ridges
and occupy topographic lows. Moreover, DEMs show that some double
ridges have an elevation that decreases in the direction of smooth
feature centers. We suggest that this could result from terrain subsidence
beneath the smooth features. Thermal erosion could also participate
to this particular morphology, but it is not required to explain the
formation of smooth features. 

To link the volume of the smooth features measured on the DEM (using
a simple approach) with the actual volume of cryolava erupted during
their emplacement, we propose to take into account a range of possible
subsidence depths, as modeled by \citet{Manga_Lenticulae_2016}. We
studied the two extreme cases of shallow subsidence of 5 m and maximum
subsidence of 40 m. In the first case, approximately 4$\times$10$^{7}$
to 4.4$\times$10$^{8}$ m$^{3}$ of cryolava are required to erupt
in order to emplace the observed smooth features. In the second case
with deeper subsidence, approximately 2.5$\times$10$^{8}$ to 2.7$\times$10$^{9}$
m$^{3}$ of cryolava are required. These volumes depend slightly on
the cryomagma composition, but this does not change significantly
the results. These extreme values may be better constrained knowing
subsidence height under the smooth features, but unfortunately, this
is not possible to infer with current data. Larger reservoirs are
expected to create deeper surface subsidence \citep{Manga_Lenticulae_2016},
so it is worth considering both of these solutions.

Using our previous model \citep{Lesage_CryomagmaAscent_2020}, we
can predict the cryolava volume erupted at the surface at the end
of a cryovolcanic eruption as a function of reservoir volume and depth.
These volumes are compared to the volumes measured on the four Galileo
images to constrain cryomagmatic reservoir volumes. We found that
a 3$\times$10$^{9}$ to 10$^{11}$ m$^{3}$ (0.9 to 2.9 km radius)
cryomagma reservoir is required to explain the emplacement of these
smooth features from a single eruption in the case of shallow subsidence
(Fig. \ref{fig:results_model_simple}, dashed lines). For deep subsidence
of 40 m, the required reservoir volumes are 10 times higher (reservoirs
up to 6 km in radius, see Fig. \ref{fig:results_model_simple}, solid
lines). In the case of cyclic eruptions from the same reservoir, for
a pure water reservoir and at the end of its activity lifetime, a
4$\times$10$^{8}$ to 4$\times$10$^{9}$ m$^{3}$ (0.5 to 1 km radius)
liquid reservoir is needed in case of shallow subsidence, and a 2$\times$10$^{9}$
to 2$\times$10$^{10}$ m$^{3}$ (0.8 to 1.6 km radius) one in case
of a 40 m subsidence. The cryomagma composition slightly changes these
results: two to three times greater reservoir volume is required for
a briny cryomagma compared to pure water (up to 2.4 km radius). The
total lifetime of such reservoirs ranges from 5$\times$10$^{3}$
to 10$^{5}$ years if we consider that eruptions stop when 100\% of
the cryomagma freezes. Nevertheless, as suggested by \citet{Quick_CryovolcanicDomes_2016},
eruptions could become impossible when 55\% of crystallinity is reached
because cryomagma viscosity at that point becomes too high. In this
case, a two times bigger reservoir is needed to explain the smooth
features observed.

The recent detection of chlorides such as NaCl \citep{Trumbo_NaClEuropa_2019}
or Mg-bearing chlorinated species \citep{Ligier_EuropaSurfaceComposition_2016}
on Europa's surface may indicate that the cryomagma composition could
be different from the one used here (81 wt\% H$_{2}$O + 16 wt\% MgSO$_{4}$
+ 3 wt\% Na$_{2}$SO$_{4}$, see \citealp{Kargel_BrineVolcanism_1991}).
Two main quantities might be impacted by the cryomagma composition:
(i) the freezing time-scale of the reservoir and (ii) the cryolava
erupted volume. The freezing time-scale is a function of the freezing
temperature of the solution, which is slightly lower for chloride
aqueous solutions than for sulfate ones \citep{Quick_HeatTransfertCryomagma_2016}.
Thus, the freezing time scale must be slightly larger for chloride
solutions. On the other hand, the erupted volume of cryomagma depends
on the density contrast between the cryomagma solution and the corresponding
ice. Here we tested a sulfate-based cryomagma and the extreme case
of pure water. The erupted volumes obtained with these two compositions
are similar for a single eruption and differ by at most a factor two
for several eruptions during the whole reservoir lifetime. Hence,
the cryomagma composition is not expected to modify the results order
of magnitude, at least for a reasonable salt content. Moreover, Eq.
(\ref{eq:Vi}) provides a very simple way to predict the cryomagma
volume erupted from a subsurface reservoir depending on cryomagma
and ice densities. It is then possible to adapt this model to any
cryomagma composition.

We demonstrated in this study that cryomagmatic reservoirs of $\sim$10$^{7}$
to 10$^{12}$ m$^{3}$ located a few kilometers under Europa's surface
may possibly be at the origin of smooth features seen on the Galileo
images 5452r, 0713r, 0739r and 9352r. This information could help
the two upcoming missions JUICE (ESA) and Europa Clipper (NASA) to
target interesting locations to search for biosignatures. In order
to better constrain the characteristic size and depth of the source
reservoirs, it would be necessary to determine whether the smooth
plains are the result of one or several eruptions. Higher resolution
images expected from future missions will provide better resolved
DEMs of the surface and more information on smooth plain morphology.
This would help to better constrain putative cryomagma reservoir dimensions.

\newpage{}

\section*{Acknowledgments}

We acknowledge support from the ``Institut National des Sciences
de l'Univers'' (INSU), the \textquotedbl Centre National de la Recherche
Scientifique\textquotedbl{} (CNRS) and \textquotedbl Centre National
d'Etudes Spatiales\textquotedbl{} (CNES) through the \textquotedbl Programme
National de Plan\'etologie\textquotedbl . We thank Benoit Jabaud
for his analysis of the Galileo/SSI images. We also thank Baptiste
Journaux for the interesting discussion we had. We gratefully acknowledge
the two anonymous reviewers for their comments and suggestions that
greatly improved this manuscript.

\newpage{}

\section*{References }

\bibliographystyle{elsarticle-harv}

\end{document}


\begin{frontmatter}{}

\title{Constraints on effusive cryovolcanic eruptions on Europa using topography
obtained from Galileo images - Supplementary materials}

\author{Elodie Lesage$^{1}$, Fr\'ed\'eric Schmidt$^{1}$, Fran\c{c}ois
Andrieu$^{1}$, H\'el\`ene Massol$^{1}$}

\address{$^{1}$ Université Paris-Saclay, CNRS, GEOPS, 91405, Orsay, France}

\end{frontmatter}{}

\section{ISIS pre-treatment \label{sec:ISIS-pre-traitement}}

The Shape from Shading (SfS) tool needs pre-processed and calibrated
images to work on. After a raw Galileo image and its label text file
are downloaded from the PDS data volumes, the first step is to make
them usable by ISIS in converting them in a .cub file, which is the
ISIS standard format. A .cub file contains the image and data from
the label text file. Then, it is necessary to add the SPICE data to
the cub file using ISIS. SPICE data contains all the information linked
to the spacecraft and imager position and orientation at the moment
of the shot. We also apply the ISIS calibration tool ``gllssical''
and the ``trim'' function which removes the noise on the outermost
pixels of the image. Finally, the image is map-projected with the
``cam2map'' tool.

\section{Uncertainties estimation \label{sec:Uncertainities-estimation}}

Here we estimate the uncertainties of the DEMs as a function of the
parameters used to generate them, in the form of a sensitivity study.
Unfortunately, absolute DEM uncertainties are not possible to calculate
on Europa, since there is no ground truth. Nevertheless, thanks to
shadows, it is possible to measure the height of some features and
use it as a comparison with the DEMs. 

From Eq. (1) in the main article, we identify two major factors that
could modify a produced DEM depending on user's choices: $\mu$ the
smoothness parameter and $R\left(h\left(x,y\right)\right)$ the reflectance
model and its parameters such as the albedo. $\lambda$, the a priori
weight, is fixed as $\lambda$ = 0 and so it has no importance on
the produced DEMs. These parameters have to be optimized and will
determine the uncertainties of a DEM. The image resolution also seems
to have an effect on a DEM, so we also tested the importance of this
bias. Finally, we generate two DEMs of the same terrain seen on two
different images to test the process repeatability. 

We choose the image 0526r (see Fig.\ref{fig:0526_zoom_ombres}) from
Galileo SSI to perform all these tests because of the heterogeneity
of the terrains and the great variety of features visible on it, which
is useful to test the parameters of the SfS tool. To test the validity
and the quality of the DEMs as functions of the parameters used, we
need to compare it with the height of the features seen on the DEMs.
To do so, the shadow measurement technique is used for each image
in this study. 

\subsection{Measurement of heights with shadows \label{subsec:shadows}}

To measure the approximate height of the reference features that we
chose, we use the shadows visible on the images. An feature of height
$h$ projects a shadow of length $l$ in the direction of the sun
lightning with: 
\begin{equation}
h=l\:\tan\left(\alpha\right)\label{eq:hauteur_trigo}
\end{equation}
where $\alpha$ is the solar elevation.

\begin{sidewaysfigure}
\centerline{\includegraphics[width=1\textwidth]{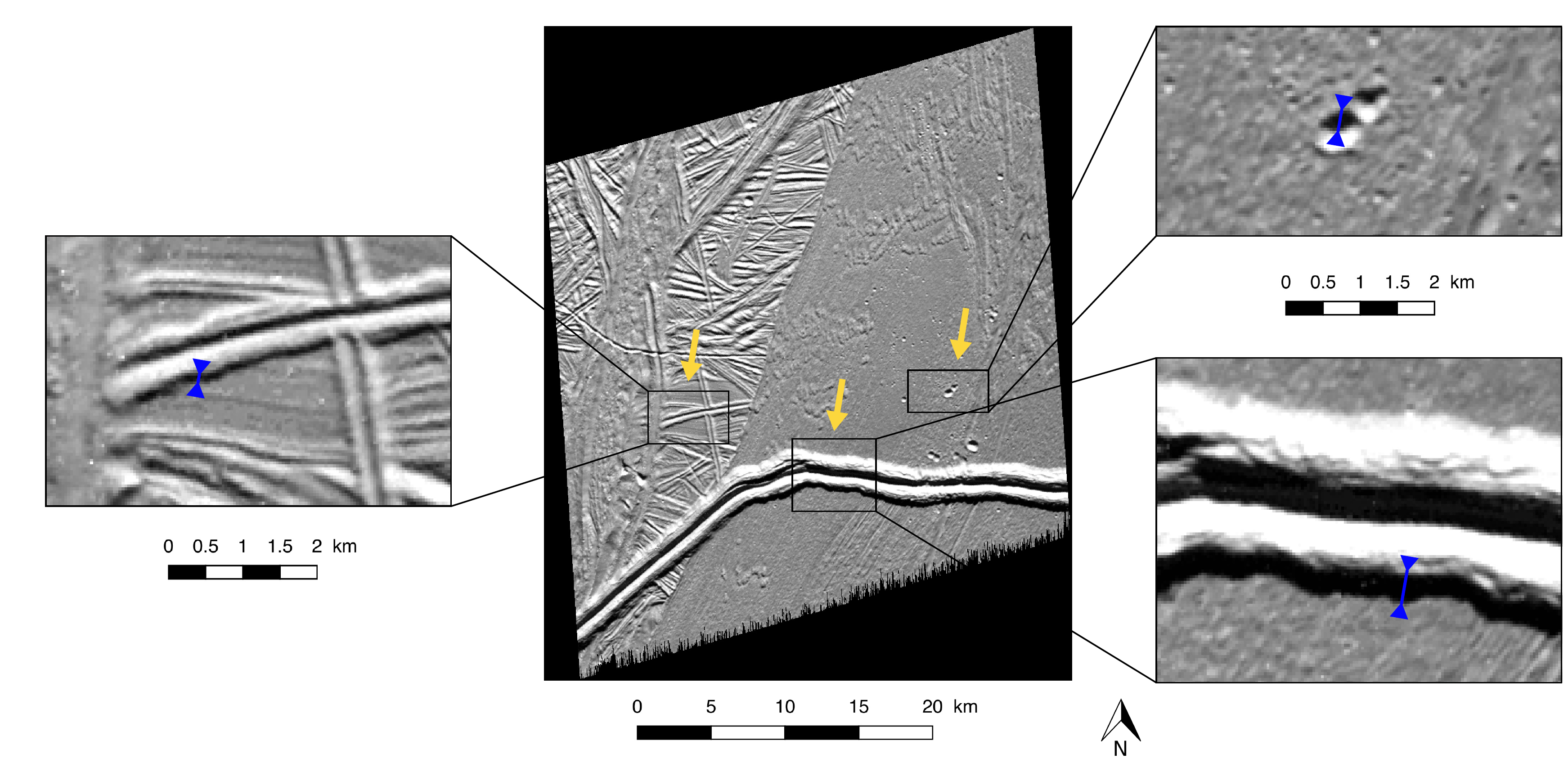}}\caption{Image 0526r from Galileo SSI. Resolution: 43 m/px. The lightning direction
is indicated by the yellow arrows. We use the shadows in the crater
and near the two double ridges shown by the blue arrows in order to
measure the height of these three features. \label{fig:0526_zoom_ombres}}
\end{sidewaysfigure}

We measured the shadow of features of three different sizes on image
0526r (see Fig. \ref{fig:0526_zoom_ombres}): the tallest double ridge
at the bottom of the image, a medium-scale crater at the right of
the image and a small double ridge at the left of the image. We obtain
a height of $170\pm30$ m for the part of the tall double ridge shown
in Fig. \ref{fig:0526_zoom_ombres}, a depth of $120\pm20$ m from
the crater rim to its center, and a height of $50\pm20$ m for the
small double ridge. These values are compared to the topographic profiles
obtained during the optimization tests detailed hereafter.

\subsection{Smoothness parameter \label{subsec:Smoothness-coefficient}}

\begin{figure}
\centerline{\includegraphics[width=1\textwidth]{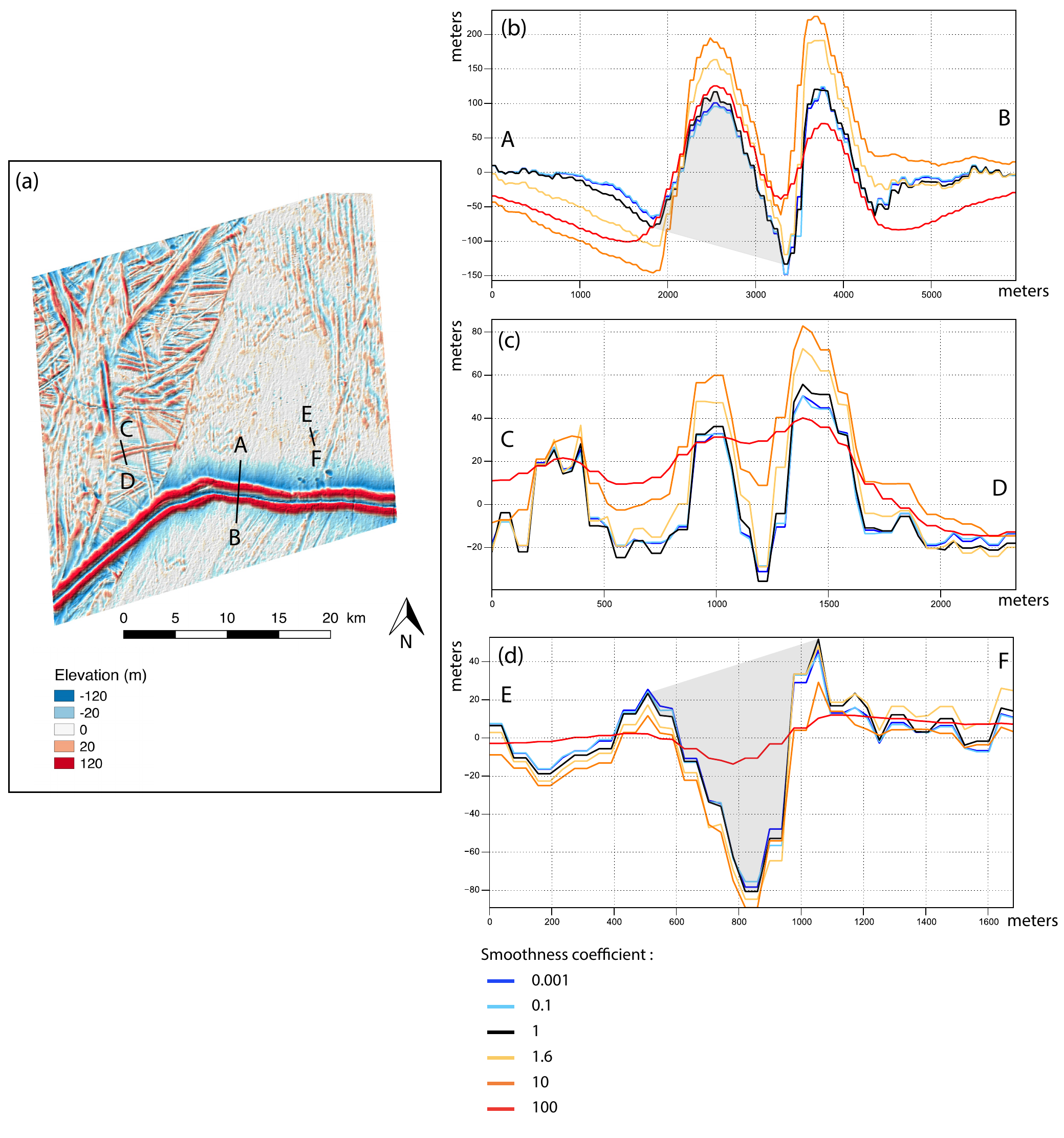}}\caption{(a) DEM of Galileo SSI image 0526r with three topographic profiles
of: (b) a high and wide double ridge, (c) a little double ridge and
(c) a medium-sized crater. The DEM has been generated using six different
values of the smoothness parameter $\mu$ ranging from $10^{-3}$
to $10^{2}$. DEM presented in (a) is calculated using $\mu=1.6$.\label{fig:coupes_smooth}}
\end{figure}

The parameter $\mu$ weights the smoothness of the final DEM. In ASP,
the default value of $\mu$ is 0.04, but the smoothness effect is
highly depending on the target surface properties \citep{Alexandrov_SfsASP_2018}
and it may be necessary to test several values of $\mu$ to find an
acceptable one, i.e. one that generates a non-noisy DEM that still
allow to show low-resolution details. To test the smoothing effect,
we computed the DEM of image 0526r using different smoothness parameters
$\mu$ ranging from $10^{-3}$ to $10^{2}$. We obtained the DEM shown
in Fig. \ref{fig:coupes_smooth}a with $\mu=1.6$. Fig. \ref{fig:coupes_smooth}b,
\ref{fig:coupes_smooth}c and \ref{fig:coupes_smooth}d are topographic
profiles of the three features selected previously obtained with various
values of $\mu$.

As expected, the DEM obtained is sensitive to the parameter $\mu$
chosen. We observe that a too high value such as $\mu\gtrsim10$ leads
to a flattened DEM at every wavelength as it is shown on Fig. \ref{fig:coupes_smooth}c
and \ref{fig:coupes_smooth}d, which is the expected effect of smoothing.
Moreover, surprisingly, a low value such as $\mu\lesssim1$ also flattens
the high-wavelength reliefs, as one can see on profile b from Fig.
\ref{fig:coupes_smooth}. Indeed, it seems that a too low value of
$\mu$ does not allow SfS to reconstruct an accurate shape of the
terrain from the slope of each pixel. To avoid these two extreme effects,
we need to use a parameter $\mu$ around 1 or 2 in the case of image
0526r. Here, we take $\mu=1.6$, a value for which we obtain the DEM
the most consistent with the heights measured previously (see sec.
\ref{subsec:shadows}). 

We also noticed that some arbitrary values of the smoothing parameter,
such as $1<\mu<1.5$ or $\mu=2$ for image 0526r, produce a very noisy
DEM, with artifacts propagating in the direction of the sunlight.
This kind of wavy noise generated by the ASP is interpreted by \citet{Jiang_ASPSfSandStereo_2017}
as an indetermination of orientation of the surface facets (i.e. the
surfaces represented by pixels) with respect to the sun direction.
Indeed, the light intensity of a facet only depends on the sunlight
incidence angle, which is not sufficient to determine the facet azimuths.
To counter this indetermination, SfS make the facet slopes consistent
with their neighbors, which most of the times gives good results.
But in some cases, because of the azimuth indetermination, once facet
orientation may be biased, and this bias then propagates across the
DEM, which generates wavy noise. We cannot explain why this effect
is produced with some specific values of the smoothness parameter,
neither predict these values. Nevertheless, we can easily detect this
kind of noise and choose another value of $\mu$ to generate our DEMs.

The several tests conducted on each image demonstrate that it is not
possible to use the same value of the smoothness parameter $\mu$
for all the images. It is necessary to adjust this value for each
image. It is also not possible to guess the most optimized value before
testing it. To produce each DEM used here, we search for a value of
$\mu$ which produces a non-noisy DEM, and which avoids the flattening
effects. To select the most appropriate value of $\mu$ for one image,
we measure the approximate height of at least two different features
using the shadow technique (see section \ref{subsec:shadows}) to
use them as reference heights. Then, we generate several DEMs with
different values of $\mu$ and compare them with the reference heights.
We keep the value of $\mu$ that generates the most consistent DEM
with the shadow measurements. The value of $\mu$ selected for each
image is given in section \ref{sec:DEMs}.

We demonstrated that the smoothness parameter affects the DEM topography,
and as in this study we focus on volume estimation, we thus need to
estimate the uncertainty of these measurements that come from the
smoothness parameter. It is not sufficient to look at the uncertainty
on the elevations since an feature could have the same volume on two
DEMs but a different mean elevation. It is thus necessary to look
at the feature cross-sections as functions of the smoothness parameter.
Fig. \ref{fig:coupes_smooth}b and \ref{fig:coupes_smooth}d, show
respectively the cross-section of the left-side of the double-ridge,
and the cross-section of the small crater, that are respectively the
greatest and the smallest features visible on this image. To infer
the uncertainty of the measured volumes, we estimate the uncertainty
on the cross-sectional areas. As an exemple, two cross-sectional areas
are shown in grey in Fig. \ref{fig:coupes_smooth}b and \ref{fig:coupes_smooth}d
and were obtained with $\mu=1$. We measure the cross-sectional areas
obtained with $\mu$ ranging from $10^{-3}$ to 10 (note that $10^{2}$
is excluded since the resulting DEM is too smooth). We then use the
maximum and minimum cross-sectional areas to calculate the percentage
of deviation of the mean, representing the uncertainty: 
\begin{equation}
\frac{\Delta CS}{CS}=\frac{\frac{CS_{max}-CS_{min}}{2}}{\frac{CS_{max}+CS_{min}}{2}}\label{eq:uncertainity_mean}
\end{equation}
where $CS$ is the cross-sectional area. To infer the uncertainty
on the volumes from the uncertainty on the cross-sectional areas,
we should also take into account the uncertainty on the feature limits
that we determine with the relation:
\begin{equation}
\frac{\Delta V}{V}=\frac{\Delta A}{A}+\frac{\Delta H}{H}
\end{equation}
where $V$ is the feature volume, $A$ is the feature area and $H$
is the feature height. $\frac{\Delta CS}{CS}$ gives the uncertainty
on the heights $\frac{\Delta H}{H}$ of the feature but does not take
into account the uncertainty on the feature area. Nevertheless, $\frac{\Delta A}{A}$
is expected to be minor compared to $\frac{\Delta H}{H}$ since the
features are sufficiently resolved on the image. Thus, we consider
that $\frac{\Delta V}{V}=\frac{\Delta CS}{CS}$.

We apply Eq. (\ref{eq:uncertainity_mean}) to the double ridge and
the small crater from the DEM of image 0526r: for the left side of
the double ridge, we measure a maximum cross-section of 239,000 m$^{2}$
for $\mu=1.6$ and a minimum cross-section of 174,000 m$^{2}$ for
$\mu=0.1$, which leads to an uncertainty of $\pm$15\%. For the small
crater, the cross-sectional area is a maximum for $\mu=1.6$ with
an area of 33,000 m$^{2}$ and a minimum for $\mu=0.1$ with an area
of 29,500 m$^{2}$, which gives an uncertainty of $\pm$5\%.

The topographic profiles in Fig. \ref{fig:coupes_smooth}b, \ref{fig:coupes_smooth}c
and \ref{fig:coupes_smooth}d show that the smoothness parameter has
a greater effect on the large features than on the small ones. This
effect is seen on every DEM we produced: the smallest features of
the DEMs have nearly the same elevation with any smoothness parameter
(as long as we choose a reasonable value). As we are interested in
small-height features, the volume uncertainty coming from the choice
of $\mu$ should not exceed $\pm$5\%.

\subsection{Reflectance model}

The SfS tool can be used with three different reflectance models,
i.e. Hapke, Lambertian, and Lunar Lambert, and for which the user
can select the parameters. The DEMs used here are generated using
the Hapke model, which is the most commonly used to model the surfaces
of icy moons \citep{Belgacem_RegionalPhotometryEuropa_2019}. Here
we test the influence of the model and parameters selected on the
resulting DEM.

\begin{figure}
\centerline{\includegraphics[width=1\textwidth]{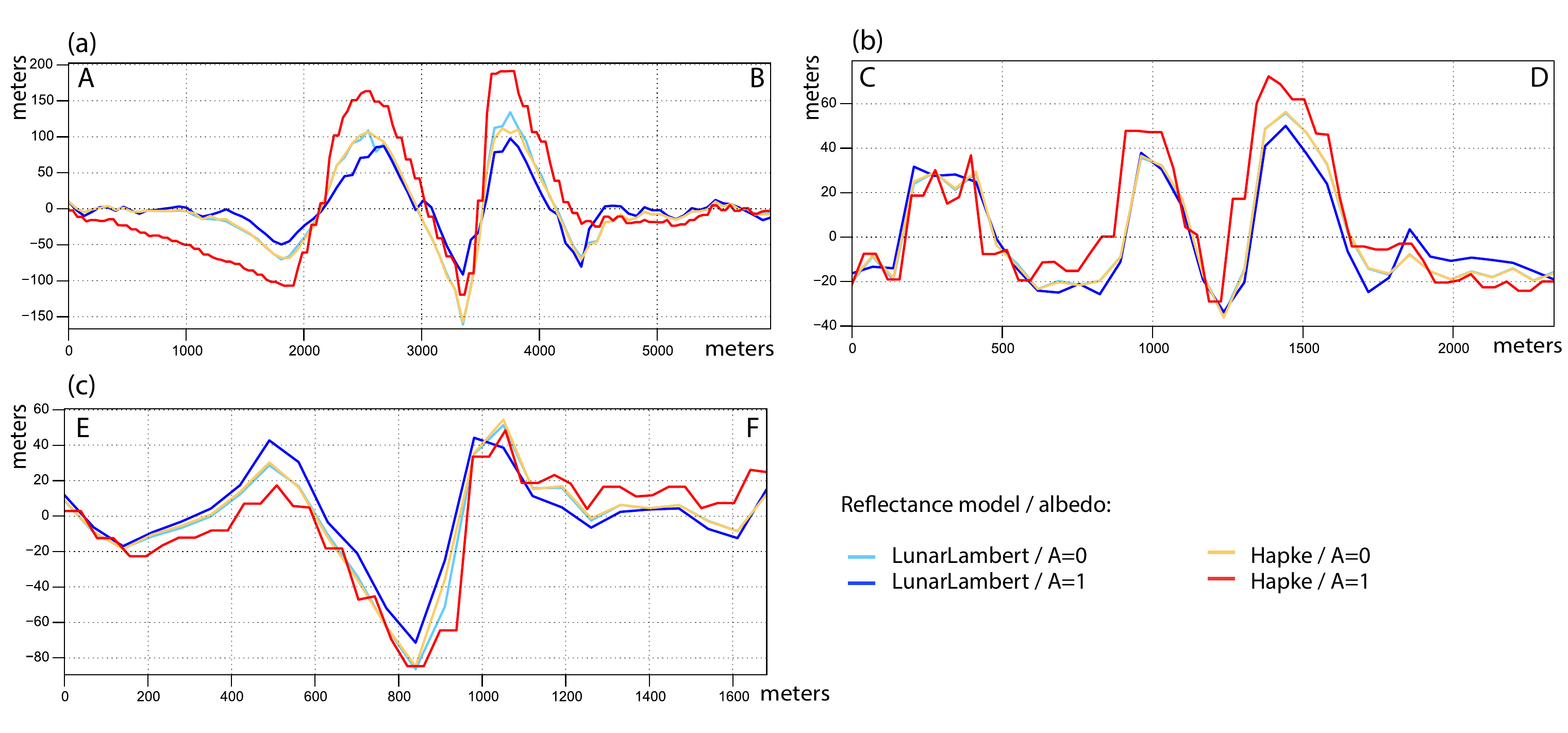}}\caption{Topographic profiles of cross-sections AB, CD and EF from DEMs of
image 0526r (see Fig. \ref{fig:coupes_smooth}) generated with two
different models (Hapke and LunarLambert) and two different albedo
parameters. \label{fig:coupes_model}}
\end{figure}

The Lambertian model is not suitable for icy surfaces because it generates
a lot of wavy noise on the DEMs \citep{Jiang_ASPSfSandStereo_2017}.
We focus on the Hapke and Lunar Lambert models. We generated the same
DEM of image 0526r in four different cases: a Hapke model with an
albedo $A$ of 0 and 1, and a Lunar Lambert model with an albedo of
0 and 1. The other parameters are left to their default values since
we did not observe a significant difference with other photometric
parameters. We compare in Fig. \ref{fig:coupes_model} the topographic
profiles obtained for cuts AB, CD, and EF (see Fig. \ref{fig:coupes_smooth}).

Surprisingly, the model and albedo used have a very minor influence
on the DEM obtained. At most, one can observe a relative difference
of a hundred meters for the tallest feature (double ridge shown on
profile Fig. \ref{fig:coupes_model}a). However, in this study, we
only measure relatively small features, and Fig. \ref{fig:coupes_model}b
and \ref{fig:coupes_model}c show that the reflectance model and the
albedo only modify the feature relative heights by a few tens of meters
at most. For this reason, we decided to choose the most commonly used
model in the literature, which is the Hapke model with an albedo of
0.9 \citep{Belgacem_RegionalPhotometryEuropa_2019}. 

The reflectance model effect must be considered in the uncertainties
calculation. As above, we calculate the uncertainty on the cross-sectional
areas on the left side of the tall double ridge, and on the small
crater (grey areas in Fig. \ref{fig:coupes_smooth}b and \ref{fig:coupes_smooth}d).
For the left side of the double ridge, we have a maximum cross-sectional
area of 239,000 m$^{2}$ for the Hapke model with the albedo parameter
$A=1$ and a minimum cross-sectional area of 121,000 m$^{2}$ for
the LunarLambert model with $A=1$, which gives an uncertainty of
$\pm$33\%. For the small crater, the cross-section is maximum for
the Hapke model with $A=0$ with an area of 33,000 m$^{2}$ and minimum
for the LunarLambert model with $A=1$ with an area of 27,000 m$^{2}$,
which gives an uncertainty of $\pm$10\%. Again, topographic profiles
from Fig. \ref{fig:coupes_model} show that small features are much
less impacted by the photometric model. We limit our study to low
elevation features so the volume uncertainty is under $\pm$10\%. 

Thus, the cumulated uncertainties on the low-elevation features due
to the smoothness parameter and photometry are, at worst, $\pm$15\%
on the volumes.

\subsection{Resolution}

\begin{figure}
\centerline{\includegraphics[width=0.8\textwidth]{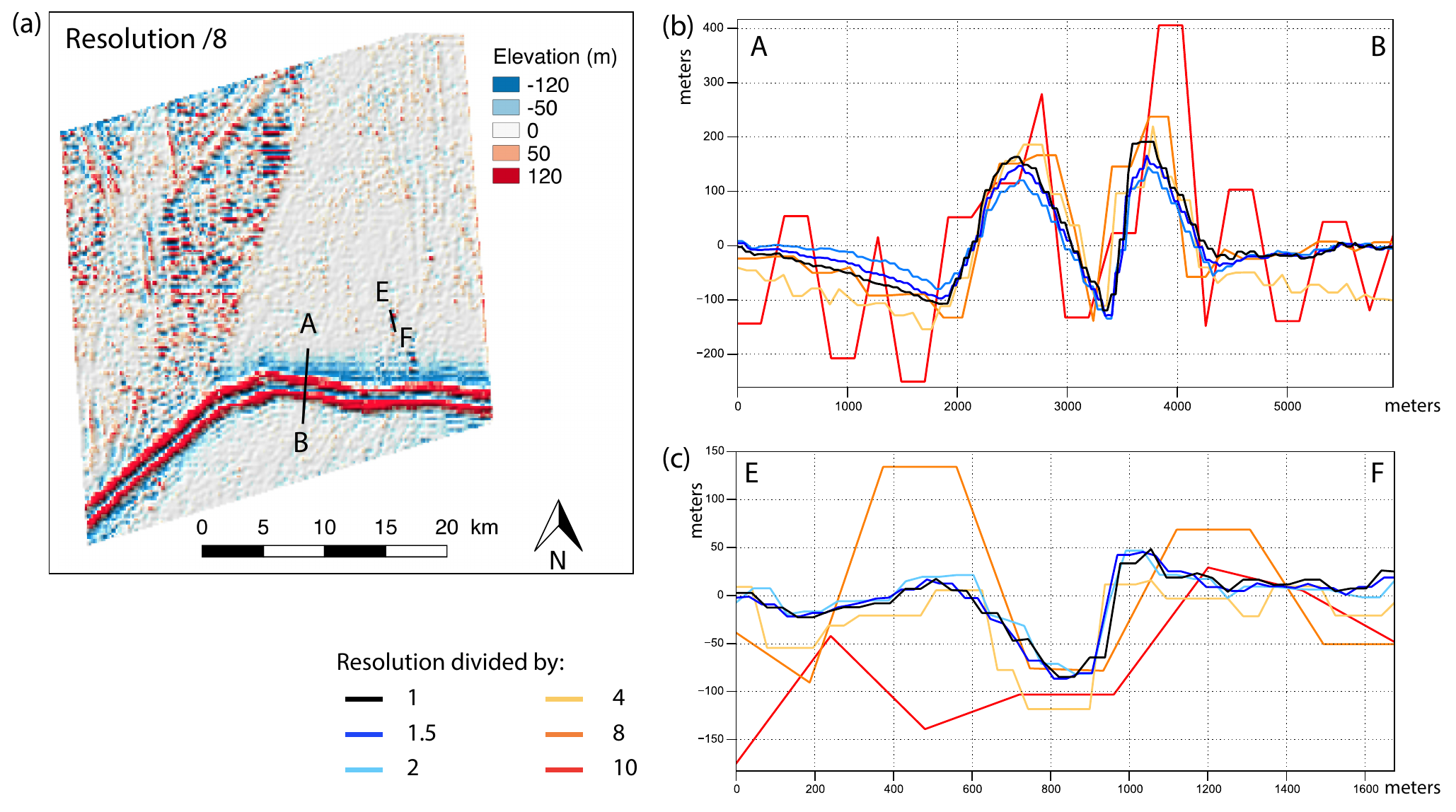}}\caption{(a) DEM generated from image 0526r that was compressed at a resolution
8 times lower than the original image. Topographic profiles of cross-sections
(b) AB and (c) EF from DEMs of image 0526r at six different resolution
compression factor.\label{fig:coupes_resolution}}
\end{figure}

Even though the image resolution is not a parameter chosen by the
user, it may have an effect on the DEM. To test this effect, we changed
the resolution of image 0526r by dividing its number of pixels. An
example is given in Fig. \ref{fig:coupes_resolution}a where the DEM
is produced with an 8 times lower resolution than the original one.
We tested resolution factors ranging from 1 to 1/10. 

We plotted the topographic profiles AB and EF (see Fig. \ref{fig:coupes_resolution}b
and \ref{fig:coupes_resolution}c) to measure the cross-sectional
areas of the tall double ridge and the small crater as functions of
the resolution factor. Globally, we noticed that the cross-sectional
area of these features becomes higher at low resolution. We found
cross-sectional areas ranging from 185,000 to 270,000 m$^{2}$ for
the left side of thetall double ridge, which gives an uncertainty
of $\pm$19\%, and from 29,000 to 67,000 m$^{2}$ for the small crater,
which gives an uncertainty of $\pm$40\%. This time, we can see that
small features are more affected by the image resolution. Nevertheless,
for reasonable resolution factors below 1/8, we can see in Fig. \ref{fig:coupes_resolution}b
and \ref{fig:coupes_resolution}c that topographic profiles are not
affected so much by resolution change: we only have an uncertainty
of $\pm$2\% on the small crater for a resolution factor between 1
and 1/4. 

\subsection{Repeatability}

\begin{figure}
\centerline{\includegraphics[width=0.8\textwidth]{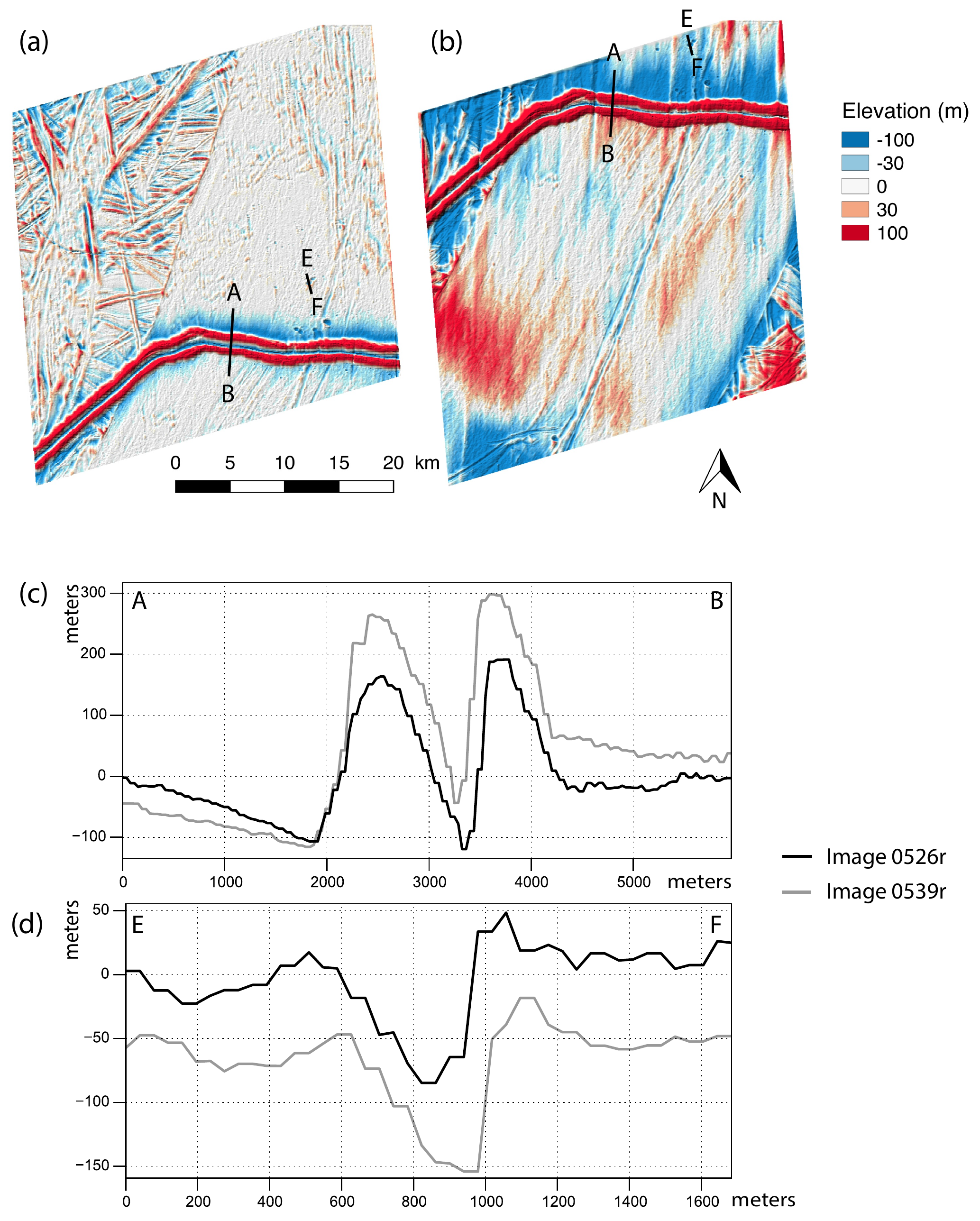}}\caption{DEMs of (a) image 0526r and (b) the overlapping image 0539r from Galileo
SSI data. Topographic profiles from these two DEMs are obtained from
cross-sections of (c) the tall double ridge and (d) the small crater.
Topography from image 0526r is plotted in black, topography from image
0539r is plotted in grey. \label{fig:coupes_coherence}}
\end{figure}

We finally test the repeatability of the process tool by generating
DEMs of two overlapping images and by comparing them. The two images
used (0526r and 0539r) were taken during the same orbit (E17), under
the exact same conditions (solar azimuth of 197°, sunlight incidence
angle of 68°). The DEMs are computed with the same input parameters:
the Hapke model with $\omega$=0.9, $b$=0.35, $c$=0.65, $B_{0}$=0.5
and $h$=0.6, the smoothness parameter $\mu$=1.6 and the initial
DEM constraint weight $\lambda$=0. In Fig. \ref{fig:coupes_coherence},
we show DEMs generated with two overlapping images 0526r and 0529r
and compare two topographic profiles from them.

Fig. \ref{fig:coupes_coherence} shows that the DEM of image 0539r
(Fig. \ref{fig:coupes_coherence}b) is much noisier than the DEM of
image 0526 (Fig. \ref{fig:coupes_coherence}a). DEM of image 0539r
presents the wavy noise evoked in section \ref{subsec:Smoothness-coefficient}
and usually avoid by selecting an appropriated smoothness parameter.
This shows that even for very similar images taken under exactly the
same conditions, the DEM production needs to be optimized by selecting
an appropriate smoothness parameter, certainly due to roughness differences
between the two images. 

We calculate the cross-sectional areas of the left side of the double
ridge and of the small crater as previously. For the double ridge,
we obtain a cross-sectional area of 253,000 m$^{2}$ for image 0526r
and a cross-sectional area of 228,000 m$^{2}$ for image 0539r, which
easily fits with a $\pm$15\% uncertainty. For the small crater, we
obtain a cross-sectional area of 27,900 m$^{2}$ for image 0526r and
a cross-sectional area of 28,300 m$^{2}$ for image 0539r, which again
is well under the uncertainties we predicted. This validates the uncertainty
range chosen. Moreover, this example confirms that $\mu$ and the
reflectance model chosen have a minor effect on small features topography.

\section{DEMs produced and volume interpolation \label{sec:DEMs}}

In this section we show the four DEMs generated with SfS and used
to measure the volume of the smooth features. 

\subsection{Image 5452r}

\begin{figure}
\centerline{\includegraphics[width=1\textwidth]{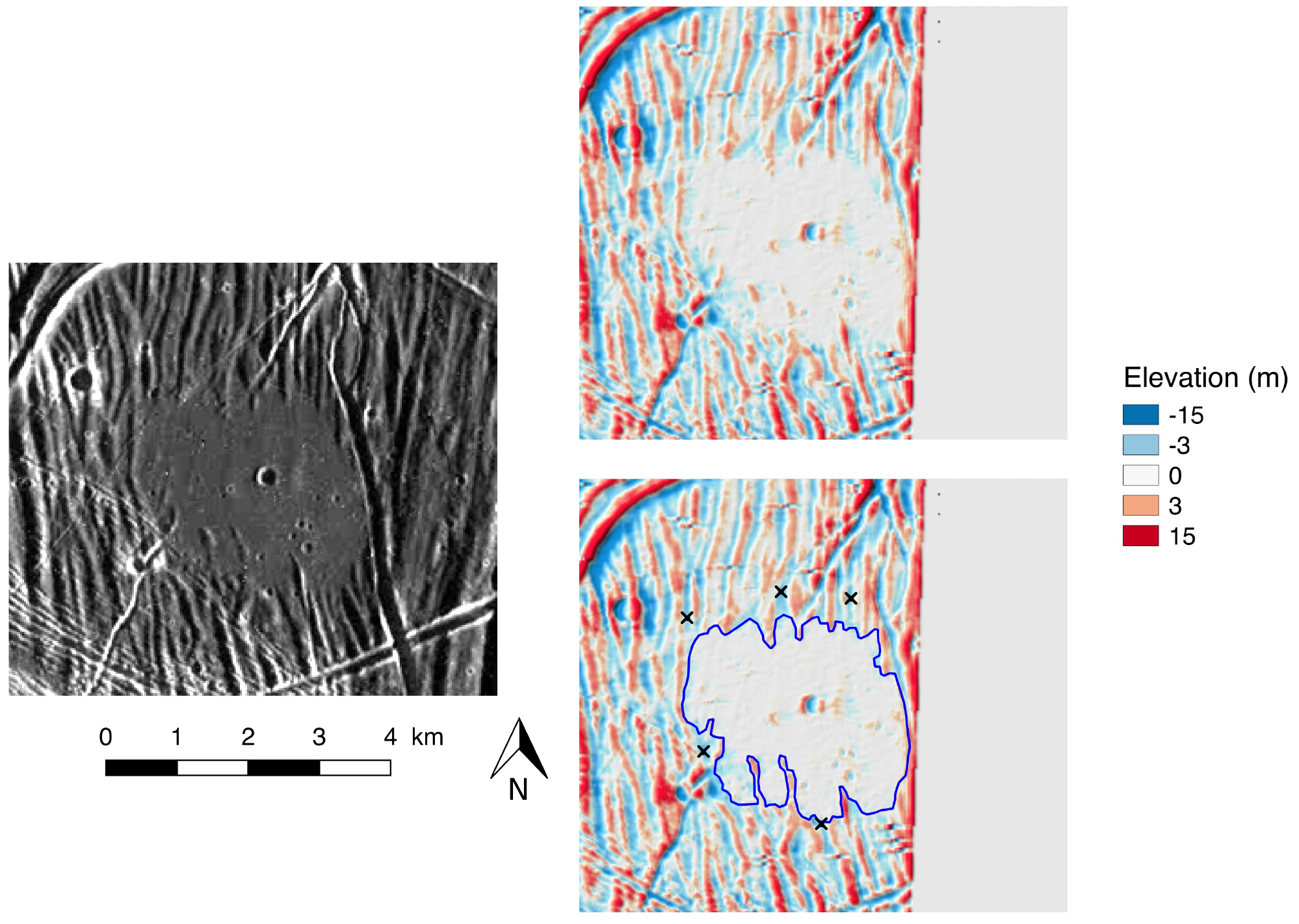}}\caption{Image 5452r from Galileo SSI (left) and the associate DEM produced
with SfS (right). Image resolution: 27 m/px. The flow-like feature
is delimited by the blue line, and the black crosses show the points
taken to interpolate a pre-existing terrain beneath the smooth feature.
\label{fig:5452r_supmat} }
\end{figure}

Image 5452r (see Fig. \ref{fig:5452r_supmat} left) is a well-known
image of Europa's surface showing a circular smooth feature. Even
if it is not obvious on the raw image, a few noisy pixels on the right
side of the smooth feature were producing noise on the smooth feature,
this is the reason why we cropped the image to generate the DEM (see
Fig. \ref{fig:5452r_supmat} left). We choose the value $\mu=0.02$
to generate this DEM. Indeed, $\mu>0.02$ leads to a noisy DEM. With
the shadow measurement, we find that the small ridge crossing the
feature to the south-east is $\sim20$ m tall, which is consistent
with the calculated DEM. The measured volume of the smooth feature
delimited by the blue line gives (5.7$\pm$ 0.9) $\times$10$^{7}$
m$^{3}$.

\subsection{Image 0713r}

\begin{figure}
\centerline{\includegraphics[width=1\textwidth]{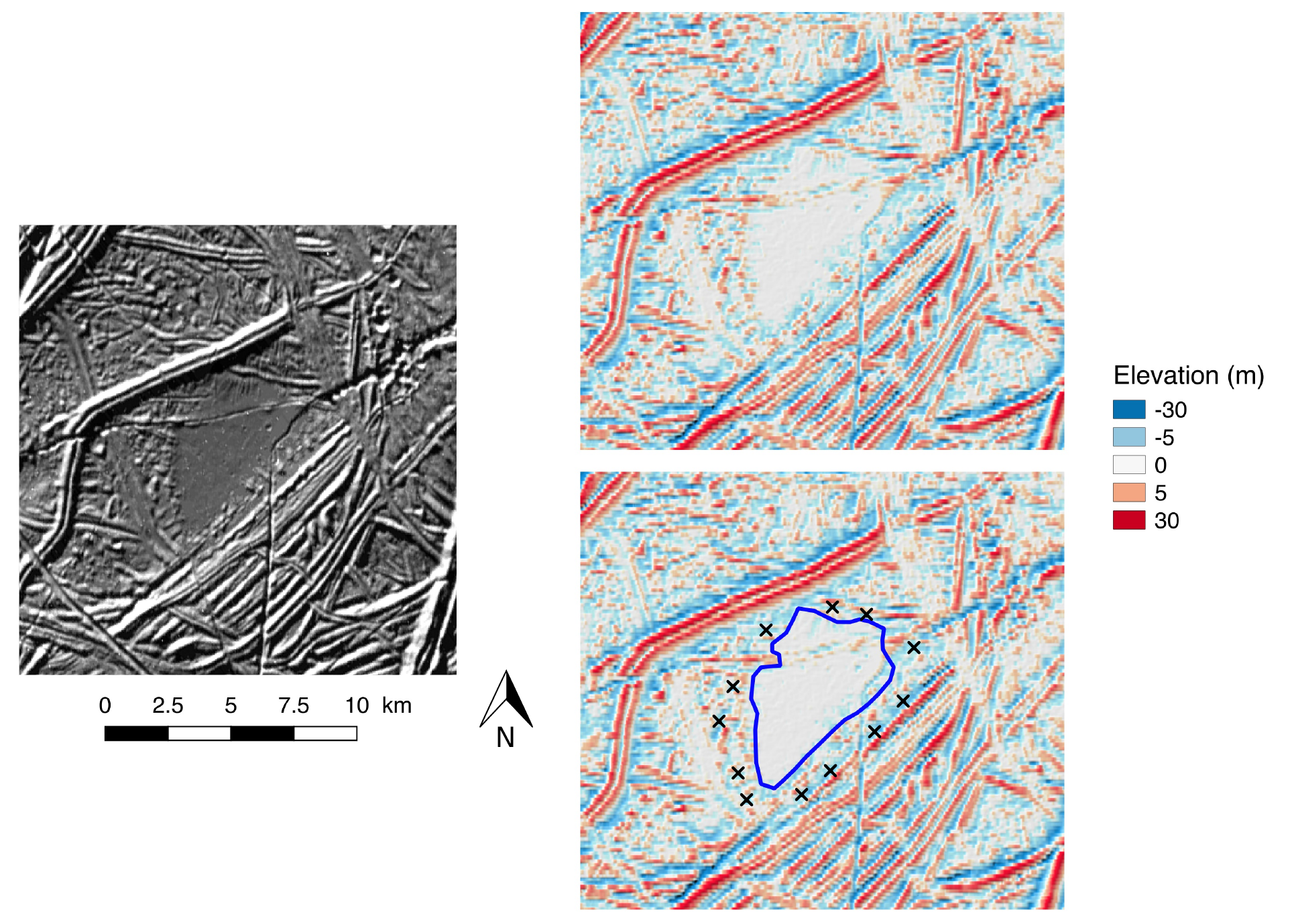}}\caption{Image 0713r from Galileo SSI (left) and the associate DEM produced
with SfS (right). Image resolution: 27 m/px. The flow-like feature
is delimited by the blue line, and the black crosses show the points
taken to interpolate the pre-existing terrain. \label{fig:0713r} }
\end{figure}

Image 0713r is shown in Fig. \ref{fig:0713r}. On this image, one
can see a smooth, relatively young feature, crossed by only two fractures.
The double ridge crossing the image north-west of the smooth feature
is found from shadow measurements to be approximately 40$\pm$10 m
tall, which is also the value found on the DEM generated with $\mu$=0.01.
We calculated the volume of the smooth feature delimited by the blue
line in Fig. \ref{fig:0713r} (right) and found a volume of (6.6$\pm$1.0)
$\times$10$^{7}$ m$^{3}$. 

\subsection{Image 9352r}

\begin{figure}
\centerline{\includegraphics[width=1\textwidth]{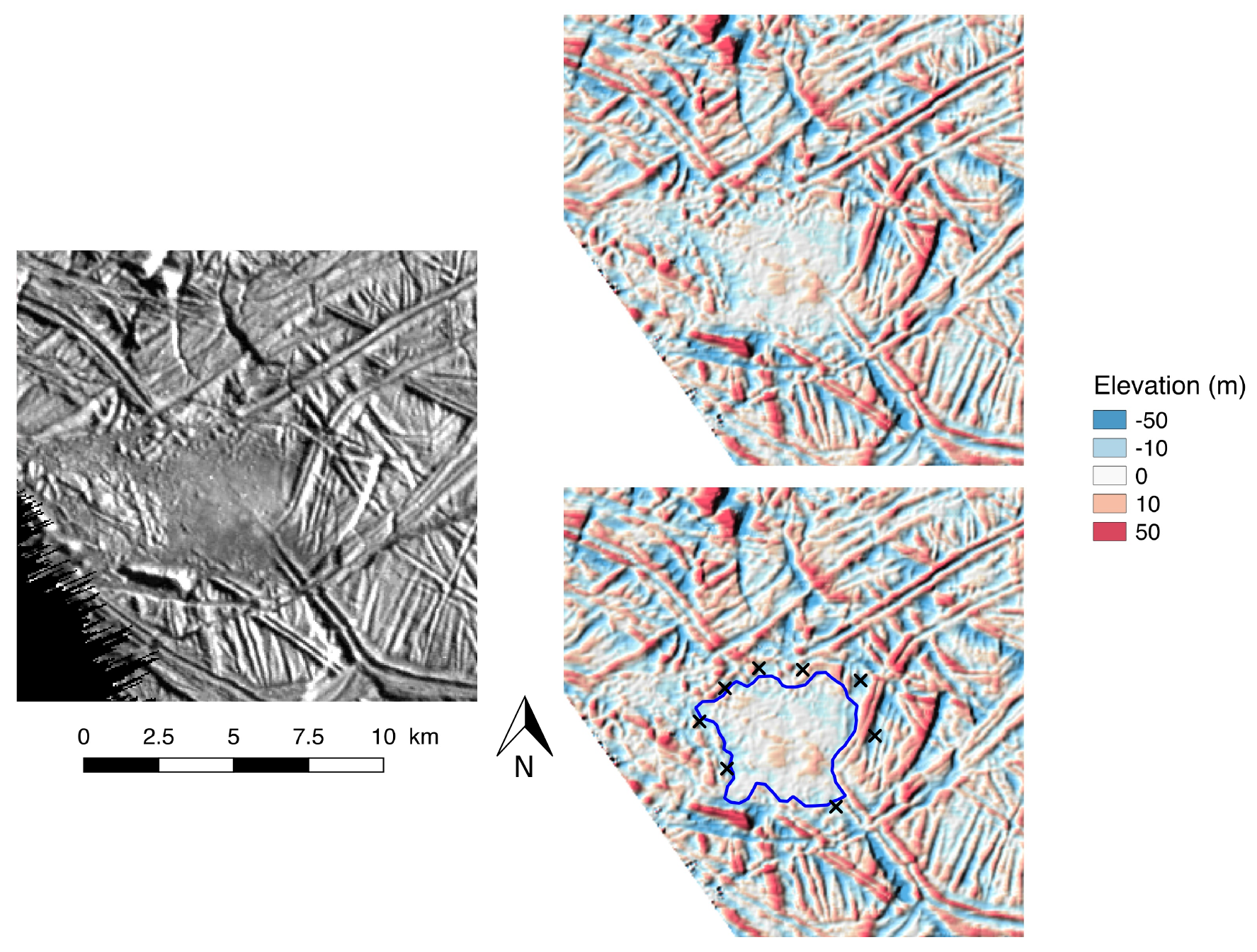}}\caption{Image 9352r from Galileo SSI (left) and the associate DEM produced
with SfS (right). Image resolution: 60 m/px. The flow-like feature
is delimited by the blue line, and the black crosses show the points
taken to interpolate the pre-existing terrain. \label{fig:9352r} }
\end{figure}

Image 9352r exhibits a smooth feature that seems younger than all
the surrounding ridges and faults (see Fig. \ref{fig:9352r}). DEM
shown in Fig. \ref{fig:9352r} is obtained with $\mu$=0.01. The high-elevation
angular feature south-west of the smooth feature is approximately
120$\pm$20 m high based on a shadow measurement, compared to $\sim$80
m high on the DEM, but the other smoothness parameters give a noisy
DEM. This should not affect the topography of small features, as shown
previously (see section \ref{sec:Uncertainities-estimation}). The
measured volume of the smooth feature delineated by the blue line
on Fig. \ref{fig:9352r} is (4.2$\pm$0.6)$\times$10$^{8}$m$^{3}$.

\subsection{Image 0739r}

\begin{figure}
\centerline{\includegraphics[width=1\textwidth]{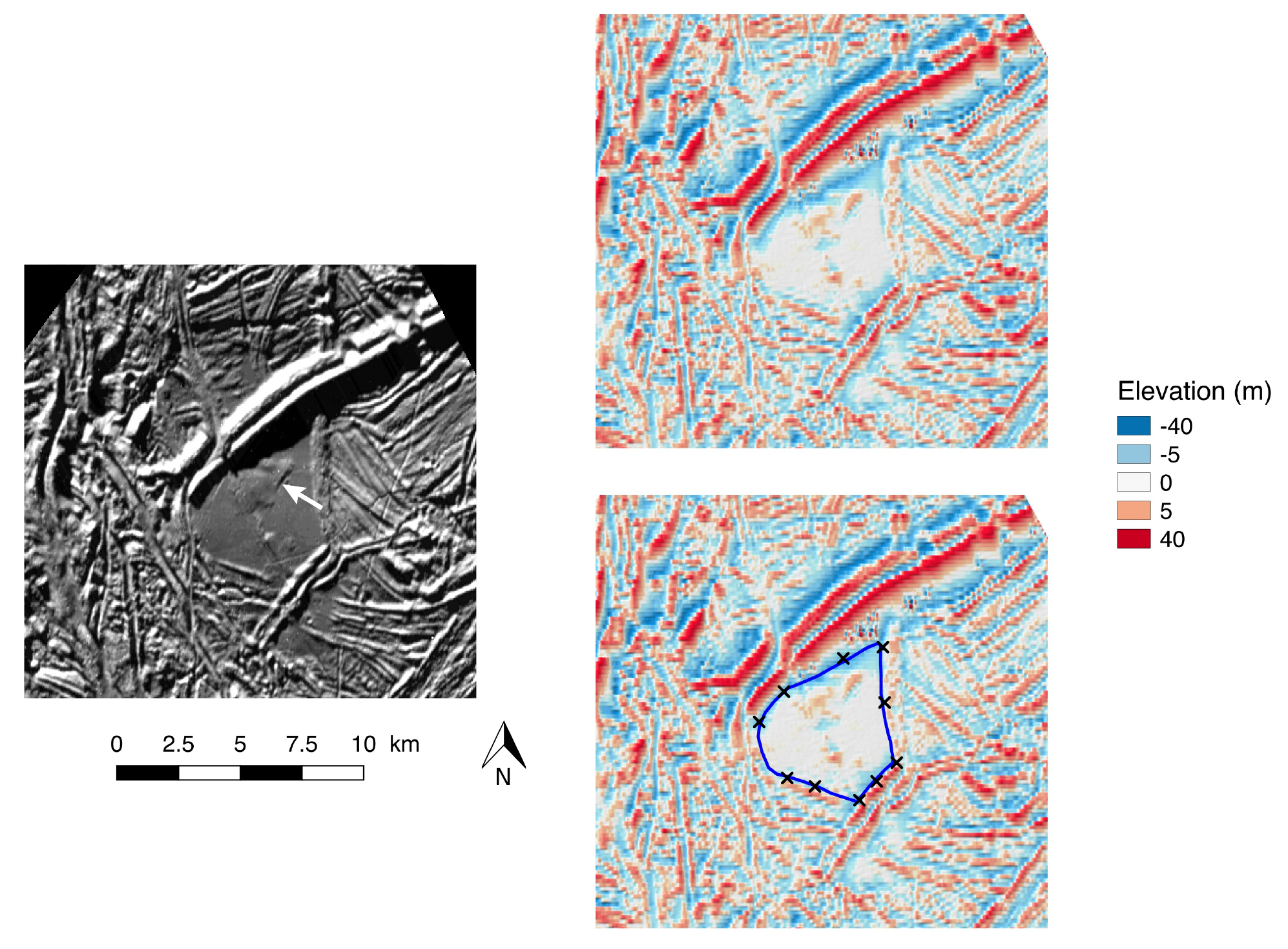}}\caption{Image 0739r from Galileo SSI (left) and the associate DEM produced
with SfS (right). Image resolution: 57 m/px. The white arrow shows
a lobate feature on the smooth feature. The flow-like feature is delimited
by the blue line, and the black crosses show the points taken to interpolate
the pre-existing terrain. \label{fig:0739r} }
\end{figure}

Image 0739r (see Fig. \ref{fig:0739r} left) shows a smooth feature,
with a small lobate elevated feature on it (see the white arrow on
Fig. \ref{fig:0739r} left). The smooth feature seems to be crossed
by several double ridges, and more precisely by a big double ridge
north-west of the feature and a small one south-east of the feature,
but it is hard to tell if the smooth material extends to the other
side of these ridges. For this reason, we only take into account the
smooth material contained between these two ridges. 

From shadow measurements, the tall double ridge north-west of the
smooth feature is measured to be 70$\pm$10 m tall, and the small
double ridge south-east of the smooth feature is 35$\pm$10 m tall.
The DEM shown in Fig. \ref{fig:0739r} (right) is generated with $\mu$=0.5.
On this DEM we can measure heights of respectively 60 and 30 m for
the tall and the small double ridges, which values are similar to
the ones measured from the shadows. We finally find a volume of (2.7$\pm$0.4)$\times$10$^{8}$
m$^{3}$ for this smooth feature.

\newpage{}

\section*{References }

\bibliographystyle{elsarticle-harv}